\documentclass[preprint]{elsarticle}
	
	\usepackage{lineno,hyperref}
	\usepackage{algorithm}
	\usepackage{algpseudocode}
	\usepackage{amssymb}
	\usepackage{listings}
	\usepackage{amsmath}
	\usepackage{svg}
   \usepackage{booktabs}
    \usepackage{multirow}
    \usepackage{makecell}
    \usepackage{lscape}
    \usepackage[flushleft]{threeparttable}
	\newdefinition{definition}{Definition}
       \journal{Journal of Systems and Software}
       \usepackage{stmaryrd}

       \usepackage{booktabs,tabularx,caption}

\begin{document}
	
	\begin{frontmatter}
	
	\title{ Runtime Verification on Abstract Finite State Models
	}
	
	
	\author[1]{KP Jevitha \corref{mycorrespondingauthor}}
	\cortext[mycorrespondingauthor]{Corresponding author}
	\ead{kp\_jevitha@cb.amrita.edu}
	
	\author[1,2]{Bharat Jayaraman \corref{mycorrespondingauthor} }
	\ead{bharat@buffalo.edu}
	
	\author[3]{M Sethumadhavan \corref{mycorrespondingauthor}}
	\ead{m\_sethu@cb.amrita.edu}

	\address[1]{Department of Computer Science and Engineering, Amrita School of Computing, Coimbatore, Amrita Vishwa Vidyapeetham, India}
	
	\address[2]{Department of Computer Science and Engineering, State University of New York at Buffalo, Buffalo, NY, USA}
	
	\address[3]{TIFAC CORE in Cyber Security, Amrita School of Computing, Coimbatore, Amrita Vishwa Vidyapeetham,India}
	
	\begin{abstract}
	Finite-state models are ubiquitous in the study of concurrent systems, especially controllers and servers that operate in a repetitive cycle.  In this paper, we show how to extract finite state models from a run of a multi-threaded Java program and carry out runtime verification of correctness properties.  These properties include data-oriented and control-oriented properties; the former express correctness conditions over the data fields of objects, while the latter are concerned with the correct flow of control among the modules of larger software.  As the extracted models can become very large for long runs, the focus of this paper is on constructing reduced models with user-defined abstraction functions that map a larger domain space to a smaller one. The abstraction functions should be chosen so that the resulting model is property preserving, i.e., proving a property on the abstract model carries over to the concrete model. The main contribution of this paper is in showing how runtime verification can be made efficient through online property checking on property-preserving abstract models.  The property specification language resembles a propositional linear temporal logic augmented with simple datatypes and operators. Classic concurrency examples and larger case studies (Multi-rotor Drone Controller, OAuth Protocol) are presented in order to demonstrate the usefulness of our proposed techniques, which are incorporated in an Eclipse plug-in for runtime visualization and verification of Java programs. 
	
	\end{abstract}
	
	
	\end{frontmatter}
	
	
	\section{Introduction}
	\label{sec:introduction}
		The context for this research is a state-of-the-art dynamic analysis and visualization environment for Java, called JIVE, for Java Interactive Visualization Environment \cite{GJ:JIVE-2005,CJ:ETX-2007,LBDSwaminathan:RV-JIVE-2016,JLSwaminathan:SPE-2017,JJJS:FSM-SPE-2021}.  JIVE enhances program comprehension by supporting multiple views of execution, through object-, sequence-, and state-diagrams, interactive forward and reverse stepping, dynamic slicing, and query-based debugging.  The diagrams resemble those of UML \cite{JBR:UML-1999}, except that they are constructed at runtime, in order to facilitate conformance checking of execution with design.  The motivation for this research stems from the fact that diagrams are useful when they are small but they can become unwieldy for long runs. Hence our interest in developing abstract models that reduce the size of diagrams as well as verification techniques on these models.
		
		The focus of this paper is on finite state models, a topic that has been extensively studied in the literature for over four decades
		\cite{Sifakis:PropPreservingHomomorphisms-1982, 
		Clarke:AutomaticVerification-FSCS-TLS-1986, Harel:StateCharts-1987,
		Clarke:ModelChecking-1997}.  In contrast with traditional model checking, which is applicable at the design stage, our focus is on the correctness of an implemented system (in Java). The term {\it runtime verification} \cite{HG:RV-2001-ase,MC:RV-Brief-2009,FHR:RV-tutorial-2013} is used to refer to a broad class of techniques that involve extracting information from an execution of a program in order to verify properties of interest. 
		As noted by \cite{HSZ-Sistla:MC+RV-2014}, the two approaches are complementary:  design-time model checking can catch errors early and can in principle explore all execution paths, while run-time verification is performed on the execution trace from actual implementation and takes into account the actual operating environment, but may not explore all execution paths.  The applications of interest to us are concurrent systems such as servers and controllers that have a cyclic operation and are non-terminating.  A single run in such systems typically exercises a large number of scenarios and execution paths.  For offline analysis, we forcibly terminate the program and obtain a finite execution trace from which finite state models are constructed.  Online analysis, on the other hand, can be carried out as the program runs and a property violation can be immediately highlighted. 
		
		Finite state models can become excessively large if the state vector includes every field of every object in the program.  Hence, our model construction method excludes all local variables of a method from the state vector -- since they are transient part of the object state -- and permits the user to selectively choose certain fields of objects, which we refer to as {\it key attributes}.  For offline analysis, we extract an execution trace that essentially records the sequence of changes to all key fields.  This sequence of state changes essentially defines a finite state model, referred to as a {\it linear state model}, the size of which is equal to the number of field update instructions in the execution trace.  This model is not constructed in practice, as it can get very large, but it is the most accurate state model for the execution trace.
		
		The main idea of this paper is that an analysis of the form of the property to be verified can help construct reduced models that cater to runtime verification of the property.  One class of properties is data-oriented, or state-based, and is of the form {\tt G}[{\it p}], meaning that every distinct state of the model should satisfy a condition {\it p}, which is a propositional formula usually extended with simple theories (numbers, strings and operators) for which validity can be checked. For this class of properties, we can construct  a {\it distinct state model} by collapsing into one state all states of the linear state model that have the same vector of values. We can take this idea further by analyzing the form of condition {\it p} and postulating {\it abstraction functions} that result in even smaller state models yet are suitable for correctly verifying {\it p}.  This criterion is often referred to as {\it property-preserving abstraction} \cite{LGSBB:PropPreservingAbstractions-1995}, meaning that a property that is verified on the abstract model also holds on the concrete model.  We elaborate on this and other approaches to abstraction in Section \ref{sec:related}.  In short, the main difference is that our abstraction techniques are applied at runtime rather than at design-time.
		
		Another class of properties is control-oriented, an important form of which is
		{\tt P}[$p\sim\sim>q\sim\sim>r$], meaning that every path from a state where {\it p} is true to a state where {\it r} is true must pass through a state where {\it q} is true. Data-oriented properties are common for intra-module/method analysis where the data fields are visible. Control-oriented properties are more common for expressing the desirable flows across modules, since higher levels of the software typically use {\it data abstraction} (i.e., data attributes are not visible) and the focus is on operations that are visible through module interfaces.
		
		Thus the main contributions of this paper lie in the construction of abstract models for both data-oriented and control-oriented properties, the definition of the property specification language with examples, and algorithms for property checking on abstract models (sections \ref{sec:abstraction}, \ref{sec:lang}, and \ref{sec:verification}). Property checking with concrete values for state-based properties of the form {\tt G}[{\it p}] requires {\it p} to be evaluated at each state by substituting each key attribute by a concrete value. For abstract models, we formulate for each state component a condition that is true for the set of concrete values that are abstracted by that state component. We then check, for each state, that the conjunction of these conditions across all components of the state vector implies the property {\it p} of interest.  For path-based properties, we construct a reduced model such that checking the path property on the reduced model is equivalent to checking the property on the linear state model.
		
		Two additional contributions of this paper are the ability to perform property checking on abstract models in an online manner (i.e., as the program runs) and also to support runtime verification for large applications and long executions. We have experimented with three substantial case studies in recent years:  a Multi-rotor Drone Controller \cite{JMAVSim1}, the Open Authorization Protocol \cite{OAuth2}, and the Apache Tomcat Server \cite{tomcatarch}.  We reported in detail on the Tomcat server in our previous paper \cite{JJJS:FSM-SPE-2021}, and hence this paper focuses on the former two case studies.  
  
  Crucial to the construction of finite state models for large case studies is the ability to efficiently generate execution events from a run of the Java program. The standard JIVE debugger is founded on JPDA, the Java Platform Debugger Architecture, which generate events by pausing and resuming the JVM (Java Virtual Machine)  during a program debug session. This pause-resume cycle is acceptable for moderate runs ($<$ 500,000 events on conventional hardware), but incurs excessive delays for longer runs.
  Hence we developed a {\it byte-code instrumentation} (BCI) technique, which bypasses the JPDA and permits the application to be executed in the `run' mode (instead of the `debug' mode) at full Java speeds. We briefly review the main ideas underlying BCI for online verification in section \ref{sec:casestudies}; a more detailed account of its performance was reported in our earlier paper
   \cite{JJJS:FSM-SPE-2021}.  
  
  This work substantially extends our previous research \cite{JJJS:FSM-SPE-2021} in showing how online verification on abstract models (along with visualization)
		can be carried out as events are generated.  We refer to the extended system as JIV$^2$E as it extends JIVE with a verification component.  
		The properties to be verified also help optimize the efficiency of execution event generation. A targeted event collection is possible through the use of an {\it inclusion filter} supported by the JIV$^2$E instrumentation module. The packages/classes/methods directly related to the properties to be verified are listed in the inclusion filter. This ensures that events are generated only for the designated units specified in the inclusion filter, thereby minimizing the overhead of event extraction.  
		
		Finally, we would like to note that although our implementation has been developed for the Java language, the ideas in this paper are applicable to other languages as well as other systems. Essentially, we require a trace of events and a meaningful notion of states and transitions. We discuss this topic further in the Section \ref{sec:conclusion} of the paper.
		
		The rest of the paper is structured as follows: Section \ref{sec:related} reviews related research in the literature and makes comparisons with our approach.  Section  \ref{sec:abstraction} presents algorithms for model abstraction and Section  \ref{sec:lang} presents the property specification language with classic examples from concurrency and case studies. Section \ref{sec:verification} describes verification of properties on abstract models and Section  \ref{sec:casestudies} presents larger case studies.  Finally, Section  \ref{sec:conclusion} presents our conclusions and directions for further work.
	
	\section{Related Work} 
	\label{sec:related}
	
	There has been considerable interest in the use of abstraction to enhance the performance of property checking of finite-state systems \cite{LGSBB:PropPreservingAbstractions-1995, CGL:MCAbst-Clarke-1994,BMMR:PredicateAbstractionCPgms-2001, BLR-Ten-SLAM, Dwyer:Bandera-Abstraction-2001}. We first briefly highlight the main approaches to design-time abstraction methods and then point out the similarities and differences with our proposed approach.

{\it Property-Preserving Abstractions.}
	 As noted earlier in Section \ref{sec:introduction}, abstractions should be property-preserving in order that 
	 verification can be carried out on abstract models.  This concept was first introduced by
	 Sifakis \cite{Sifakis:PropPreservingHomomorphisms-1982} in the context of transition systems and further studied by Loiseaux, et al
 	\cite{LGSBB:PropPreservingAbstractions-1995}.
 They show the preservation of properties in a fragment of the $\mu$-calculus when the states of the two transition systems are related by a Galois connection.  This is a specialization of the  approach by Cousot and Cousot
 \cite{CC:AbstractInterpretation-1977, CC:AbstractInterpretation-1979} who showed the use of Galois connections for a language-independent way of specifying abstract interpretations of programs. Abstract interpretations  \cite{CC:AbstractInterpretation-1977, CC:AbstractInterpretation-1979} are generally used to collect information about the static semantics of a program, in contrast with our work which focuses on the dynamic behavior of programs. In abstract interpretations, the abstractions are defined for particular types of analysis and are fixed, whereas, in our approach, the abstractions are user-defined and given on the fly.
	
	Clarke et al \cite{CGL:MCAbst-Clarke-1994} showed the use of abstraction to reduce the complexity of checking properties in a CTL branching-time logic and to facilitate the verification of problems with very large state spaces. The transition system is defined using a finite-state procedural program and symbolic execution is performed on this program in order to produce an abstract representation of its behaviour. 
	The temporal properties are then verified using state-space search in the abstract space.
	Property preservation holds for abstraction in the sense that if a property is true in the abstract space it is true in the original space; if it is false in abstract space, it may or may not be false in the original space.
	The abstraction functions discussed in the work are: congruence modulo integer for arithmetic operations; single-bit abstractions for bit-wise logical operations; product abstractions for combining various abstractions;
	and symbolic abstractions.  

{\it Software Model Checking.}
	In software model checking, the state space could be infinite and abstraction is essential for reducing the verification of an infinite-state system to that of a finite-state system. Ball et al \cite{BMMR:PredicateAbstractionCPgms-2001}, \cite{BLR-Ten-SLAM}   
	introduce the concept of {\it predicate abstraction} whereby a C program is abstracted as a boolean program with the aid of a set of predicates. The abstracted program has the same control flow as the C program and has been used for various static analyses as well as to check temporal safety properties of Microsoft's device drivers. The authors note that the drivers are essentially control-oriented programs and are amenable to model checking by accurately abstracting the flow of control.  A noteworthy aspect of their system (SLAM) is the ability to automatically find abstraction predicates as well as to modify them when a particular choice of predicates is insufficient to prove or disprove a property.
	
	The Bandera system developed by Dwyer et al \cite{Dwyer:Bandera-Abstraction-2001} shares with our work the focus on Java programs (or JVM bytecodes) and support for similar types of user-defined abstractions that are tailored to the temporal property being verified.  Like SLAM, the analyses are done statically, starting from a transformation of a concrete (Java) program to an abstract program. The temporal LTL property is also mapped to a property in the abstract space and the abstract program is then shown to satisfy the abstract property. Towards improving overall performance, Bandera removes irrelevant parts of the program from being considered by using program slicing based upon information about variables present in the property to be proved as well as other variables upon which they are control- or data-dependent. The main difference between the above approaches and this paper is our focus on runtime, as opposed to design-time, verification. 

{\it Specification Mining.} The concept of model abstraction has also been explored by the state-based specification mining \cite{DLWZ:MiningObjectBehaviourADABU-2006,MTR:StateBasedTesting-AJAX-2008, MMNTB:Revolution-AutomaticEvolutionMinedSpecs-2012} and monitoring \cite{ODDL:ControlledBurstRecording-CBR-2020} approaches. In object behavior mining \cite{DLWZ:MiningObjectBehaviourADABU-2006}, the program executions are observed to construct models of a Java object and these models are subject to fixed abstraction patterns to create an abstract model of the object. JIV$^2$E is similar in terms of using program execution to extract models and applying  abstractions to reduce the states, but differs in extracting models of the entire program under execution rather than focusing on a single object behavior. Also, the abstractions are defined by the users based on the domain of the key attributes selected for model construction.
 
 Marchetto et al \cite{MTR:StateBasedTesting-AJAX-2008} used state abstractions to abstract state machines of Ajax based web applications.  
 While their abstractions are similar to boolean abstractions, the JIV$^2$E user can also define 
 multi-valued and control abstractions.  Mariani et al \cite{MMNTB:Revolution-AutomaticEvolutionMinedSpecs-2012} explore the application of state abstractions to automatically evolve mined specifications from Java \cite{DLWZ:MiningObjectBehaviourADABU-2006} and Ajax \cite{MTR:StateBasedTesting-AJAX-2008} applications. The traces from the applications are analyzed to keep the  existing models updated and aligned with the actual implementations, which dynamically evolve due to changes in configuration or dynamic installation/removal of components. 

Unlike this technique, JIV$^2$E constructs the runtime model by adding new states (concrete / abstract) to the model being constructed as the events are received. The states once added are not removed from the model, thereby retaining the history of the application state. Cornejo et al \cite{ODDL:ControlledBurstRecording-CBR-2020} explore the use of automatically derived abstraction functions using symbolic execution techniques.
 
 While the derived abstractions are based on path conditions over state variables and program inputs, 
 JIV$^2$E supports user-defined abstractions such as boolean, multi-valued, control and path abstractions.
	
{\it Object Constraint Language}.
    There are some similarities between our property specification language and the Object Constraint Language (OCL) \cite{OCL}. 
    OCL is a textual language that augments UML with design-time specifications, notably constraints and rules on UML models that cannot be captured by the structural diagrams. Other key design-time specifications of OCL are pre- and post-conditions for methods and class invariants. These are a complementary set of concerns compared with the focus of this paper, which is on run-time verification of data- and control-oriented properties. The latter are typically path properties that are especially useful in larger applications.  Although there is some research on defining temporal extensions to OCL \cite{PM-2003-OCL,KT-2012}, this is still not an official part of OCL. A comprehensive verification system should support both design-time and run-time verification. The focus of this paper, however, is on the latter, and, in particular, the use of abstractions to obtain smaller models for verification. 

  {\it Runtime Verification.}
	  The early forerunners in the area of runtime verification were the monitoring and checking framework (MaC) \cite{IBKKSV:MaC-1998,KVKIS:Java-MaC-FMSD-2004} and the Java path explorer (JPAX) \cite{HG:RV-2001-ase,HG:RV-JPAX-2001,HG:RV-JPAX-2004}. Our online verification system, JIV$^2$E, is related to the MaC architecture \cite{IBKKSV:MaC-1998,KVKIS:Java-MaC-FMSD-2004} for continuous monitoring and checking of correct behaviors of an application. The MaC system is implemented for Java and the instrumentation is performed at the byte-code level.  It consists of three main components (filter, event recognizer, and runtime checker), which effectively map  low-level information, such as the values of program variables, to abstract events that serve as a basis for checking the acceptance of runtime behaviors.  JIV$^2$E also employs the above three components, but a major difference in our approach is that we directly construct an abstract model as the program runs and also carry out property checking directly on the abstract model. 
	  JPAX tests execution traces of Java programs against logic-based specifications using {\it logic-based monitoring} and detects concurrency errors, such as deadlocks and data races, using {\it error pattern analysis}. In JPAX, the logic specifications and the error pattern analysis are performed directly on the execution trace, whereas in our work, abstract runtime models are constructed before property checking. 
	  
	  While the earlier systems were designed to handle propositional events (not carrying data), the more recent systems also handled parameterized events (events with data values) \cite{BGHS:EAGLE-2004,MarceloHavelund:RV-Java-HAWK-2005,Allan:TRACEMATCHES-AspectJ-2005,StolzBodden:JLO-AspectJ-2006}. Another approach that supports both types of events is the paradigm of {\it monitor-oriented programming} (MOP) \cite{CG:MOP-2003, CMG:MOP-2004,FG:MOP-2007,CG:MOP-2009, MJGG:MOP-2012}. In MOP, the monitors are automatically synthesized from high level specifications and take appropriate actions in the event of violation or validation of properties. JIV$^2$E also supports propositional and parameterized events which we classify as control-oriented and data-oriented respectively. An important feature of JIV$^2$E is that the values of fields of objects can also be included in the finite state model, and these fields can be subject to abstraction\footnote{Although our focus is on fields of objects through FieldWrite events, we can also access method arguments through VariableWrite events}.

 The bytecode instrumentation technique used in JIV$^2$E for emitting events may be
	 contrasted with approaches based upon aspect-oriented programming (AOP)\cite{Kiczales:AOP-1997} such as DiSL \cite{Binder:DiSL-dynpgmAnys-SCP-2015} 
	 which require the source code modifications to express instrumentation details using AOP's pointcut/advice model.  Our system can emit various types of events such as {\it field write}, {\it method calls}, {\it method returns}, etc., whereas DiSL supports only events based on {\it methods calls} and its arguments, but does not allow one to extract details regarding {\it field writes} in a straightforward manner \cite{Binder:DiSL-dynpgmAnys-SCP-2015, GGANJ-Disl-2023}.
  
 Finally, we would like to note that this paper is a continuation of our recent paper \cite{JJJS:FSM-SPE-2021} whose focus was on model extraction from an execution trace derived from a run of a Java program.  Although there was some discussion of abstraction and properties in our earlier paper, we present here a more in-depth account of these topics and, in particular, provide algorithms for property checking on abstract models along with case studies of their applicability. We construct data-oriented models and control-oriented models according to whether the property to be verified is about the data states or the flow of control, respectively. The abstraction functions can be boolean or multi-valued for data-oriented models and they can be package/class/method/thread and path abstractions for control-oriented models. The property specification language caters to state-based as well as path-based properties.  We show that property preservation holds in that if the property is true in the abstract model then it is true in the concrete model. If a property fails in the abstract model, it may or may not fail in the concrete model -- the chosen abstraction may be incorrect and may need to be reformulated.
 
	\section{Construction of Abstract Models} \label{sec:abstraction}

	 A finite state model is essentially a directed graph of nodes and edges.  Each node contains a vector of values, called the {\it state vector}, which represents the state of a system at a particular point of execution. The state vector components correspond to the values of a subset of the fields, or attributes, called {\it key attributes}. Different combinations of key attributes yield different finite state models (or views) of the execution.  
	
	\begin{definition}
		A {\it key attribute} is of the form {\it p.c:i.f}, where {\it p.c:i} is the $i^{th}$ object-instance of class $c$ in  package $p$, and {\it f} is a primitive field, such as integer or boolean, or a String in class $c$. 
	\end{definition}
	
	It is easy to see that if the fields of all objects and all local variables of each method invocation are included in the state vector, the resulting  finite state model can become large and unwieldy.  Note that, even a simple integer-valued field taking a large number of values during execution could lead to {\it state explosion} -- this is where abstraction is desirable to obtain manageable models.  Hence, we permit the user to selectively include the fields of objects which form the {\it key attributes}. We exclude local variables of method calls from the state vector as they form a transient part of an object's state. 
    To further reduce the size of the state model, the user may specify that the individual object-instances of a class must not be distinguished for any chosen key attribute {\it f}. The specification of the key attribute effectively becomes {\it p.c.f}, i.e., without the object-instance indicator {\it i}. Thus the values assigned to key attribute {\it f} across the different object-instances are considered as assigned to a single attribute {\it f} in the state-vector.
	
	We may distinguish two broad categories of finite state models: {\it data-oriented} and {\it control-oriented}. Data-oriented models are generally used to make in-depth analyses based on the data fields of classes/packages, whereas control-oriented models help with the flow of control through methods/classes/packages/threads and are useful to get an overview of the working of large applications. 
	In data-oriented models, the state vector is made up of the data fields of objects, whereas in control-oriented  models the state vector holds control information, such as an instruction pointer or a package/class/method name which is being executed. We can also design hybrid models, where the state vector can contain both data and control information.

Model construction is based upon an {\it execution trace}  and a set of chosen  {\it key attributes}.  We first present
a {\it linear state model} and the {\it distinct state model} in subsection \ref{subsec:LSM-Abstraction} and this is followed by model construction for data-oriented models in subsection \ref{subsec:UserAbstraction}
and control-oriented models in subsection \ref{subsec:control-oriented}.
These models differ in size and scope for a given combination of execution trace,  key attributes and the abstraction criteria. 
	
	\subsection{Linear and Distinct State Models}
	\label{subsec:LSM-Abstraction}
The linear state model (LSM)  is the most accurate model capturing the state changes due to assignments to {\it key attributes}. These assignments are captured in {\it FieldWrite} events in the execution trace \cite{JJJS:FSM-SPE-2021}. The term `linear' means that every state, except the last, has a unique next state and there is no loop in the state graph.   
	
\begin{definition}
	Given a set of {\it key attributes} $\langle x_1,...,x_m \rangle$, a state is of the form $\langle x_1=v_1,...,x_m=v_m \rangle$, where $v_1,...,v_m$ are the assigned values to the components of the state vector.  The initial state is undefined, denoted as $\langle x_1=?,...,x_m=? \rangle$. 
\end{definition}
	
\begin{definition}
		Given a state $s = \langle x_1=v_1,...,x_m=v_m \rangle$ and a {\it FieldWrite} event $x_k \leftarrow w$, where $1\leq k\leq m$, the new state $s' = s[x_k \leftarrow w]$ and we define $s[x_k\leftarrow w]$ =
		$\langle x_1=v_1,...,x_{k-1}=v_{k-1}, x_k = w, x_{k+1} = v_{k+1},...,x_m = v_m \rangle$.
\end{definition}

	\begin{algorithm}[H]
		\caption{Linear State Model Construction Algorithm}
		\label{alg:LSM}
		\textbf{Input:} {Execution trace $T[1..n]$, Key attributes \{$x_1,\cdots,x_m$\}}\\
		\textbf{Output:} {Linear State Transition Graph $G(V,E)$}
		\begin{algorithmic} [1]
			\State $s_0$ = $\langle x_1=?,\cdots,x_m=? \rangle$;
			\State $V$  $\gets$  $[s_0]$;   $E \leftarrow [~];$
			\State $s$ $\gets$ $s_0$;
			\State {\textbf{for each} {\textit {FieldWrite Event} $x \leftarrow w$ in $T$ }} {\bf do}
		
			\State \quad {\bf if} { $x$ $\not\in$ \{$x_1,\cdots,x_m$\} } {\bf then}
			\State \qquad {\bf continue};
		
		    \State  \quad $s'$ $\gets$ $s$[$x \gets w$]; 
		    \State  \quad $V$  $\gets$ $V$ + [$s'$];  	
		    \State  \quad $E$ $\gets$ $E$ + [$\langle$$s$, $s'$$\rangle$]; 
			\State  \quad $s$ $\gets$ $s'$;
			\State {\bf end for}
			\State {\bf return} $G$;
		\end{algorithmic}
	\end{algorithm}
	
\begin{figure}[tbh]
		\includegraphics[width=\textwidth]{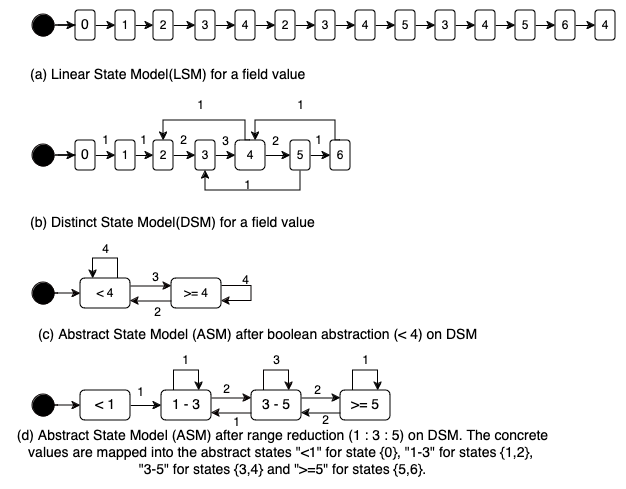}
		\caption{Abstractions for data-oriented models.}
		\label{fig:lsm-dsm-asm}
	\end{figure}
	
	Algorithm \ref{alg:LSM} shows how an LSM is constructed from an execution trace and a set of key attributes.
	In creating a new state $s' = s[x_k \leftarrow w]$, the value for $x_k$ is set to $w$ in $s'$ and all other attributes from $s$ are left unchanged. The newly created state is added to the list of vertices (V) and the transition to this state is added to the list of edges (E). Figure \ref{fig:lsm-dsm-asm}a illustrates an LSM constructed for a single integer-valued key attribute $k$ with values in the range 0..6. 
	
	The number of states created in the LSM is equal to the number of {\it FieldWrite} events on the key attributes.  Thus the size of an LSM can become very large for long executions and this motivates our interest in the construction of reduced models.
	
	\begin{definition}
	 Given an execution trace T and a set of key attributes A, the {\it distinct state model} based on T and A is obtained  by replacing in Algorithm \ref{alg:LSM} the {\it lists} of vertices and edges by {\it sets} of vertices and edges, respectively, and by replacing the {\it list append} (+) operation by {\it set union} (U). 
	\end{definition}
	  
	  \begin{definition}
	  Given an execution trace T, a set of key attributes A, and a distinct state model D based on T and A.  The {\it transition count} associated with a transition $\langle s,t \rangle$ in model D is equal to the number of transitions of the form $\langle s,t \rangle$ in the {\it linear state model} L based on T and A.
	  \end{definition}
	  
	  A distinct state model D has at most one state for any given state vector of values. A cycle will be present in the distinct state model D, if some state repeats in its corresponding linear state model L.
	 
	  Figure \ref{fig:lsm-dsm-asm}b shows the DSM constructed for the LSM shown in Figure \ref{fig:lsm-dsm-asm}a. Note that, although the number of states has reduced, the paths between states has increased due to the merging of repeated states. For example, comparing Figures \ref{fig:lsm-dsm-asm}a and \ref{fig:lsm-dsm-asm}b, there is no path from state 6 to 3 in the LSM, but there is a path from state 6 to 3 in DSM, via the states 4 and 5.
	  Thus, the DSM is suitable for checking state-based properties but not for path-based properties.

	\subsection{Abstractions for Data-oriented Models}
	\label{subsec:UserAbstraction}
	
Data-oriented models capture the data states of the application and hence we define an {\it abstract state model} (ASM) starting from an execution trace, a set of key attributes, and a set of abstraction functions, one for each component of the state vector. An abstraction function maps a larger domain of values to a smaller one and is defined by the user taking into account the properties to be verified. One abstraction function is defined per component of the state vector -- the identity function is used when abstraction is not required for some component.
 Essentially, during model construction, for every {\it FieldWrite} event on a key attribute, $x \leftarrow w$, 
 the value {\it f}$(w)$ is inserted into the state component corresponding to $x$, where {\it f} is the applicable abstraction function.
		 
Next we describe two\footnote{In our earlier paper \cite{JJJS:FSM-SPE-2021} we also introduced {\it subgraph abstraction}, but, as this is less common and does not arise in the examples in this paper, we omit its discussion in this paper.} types of abstractions for data-oriented models:
    {\it boolean abstraction} and  {\it multi-valued abstraction}.  We will discuss abstractions for control-oriented models in section \ref{subsec:control-oriented}.
	 
	 \begin{enumerate}
		
		\item
		{\it Boolean abstraction}.  Here, the abstraction function $f_i$ for attribute $x_i$ maps the domain of $x_i$ to a boolean value.  That is, $f_i$ is a predicate and the supported predicates in our implementation for the numeric attributes are {\it x} $\not=$ {\it c}, {\it x} $<$ {\it c}, {\it x} {\tt =} {\it c}  and {\it x} $>$ {\it c}, for some specific constant {\it c}. For other datatypes such as booleans, chars, enumerations and strings, only $=$ and $\not=$ are supported. Thus, the data domain of key attribute $x_i$ is reduced from a set of size {\it n} in the original model to a set with two values in the abstract model. 
		
		\item
		{\it Multi-valued abstraction}. A generalization of boolean abstraction is multi-valued abstraction, a common form of which is {\it range abstraction}, specified as $[c_1:c_2]$, which has three abstract values, corresponding to values less than {\it c}$_1$, values in the range  {\it c}$_1$ to ({\it c}$_2-1$), and values greater than or equal to {\it c}$_2$. This is a ternary abstraction and can be generalized to a multi-valued range abstraction written as $[c_1:c_2:\dots:c_n]$. 
	\end{enumerate}
	
Algorithm \ref{alg:ModelAbstraction} shows how an abstract data-oriented model G($V^a,E^a$) is constructed from an execution trace, a set of key attributes, and a set of abstraction functions $f_1\dots f_m$. 
Figure \ref{fig:lsm-dsm-asm}c shows the construction of an abstract state model for the {\it boolean abstraction} ($k<4$) on the key attribute $k$ of the model.  This results in two abstract states corresponding to ($k<4$) and ($k>=4$). Similarly, Figure \ref{fig:lsm-dsm-asm}d shows the result of applying a {\it multi-valued abstraction} range abstraction (1:3:5), which results in four abstract states in the model.

	\subsection{Abstractions for Control-oriented Models}
    \label{subsec:control-oriented}
    
	Control-oriented models deal with the flow of control among packages, classes, and methods at runtime. These models are founded  on a control variable depicting the current location of execution inside the application. The location information can be coarse-grained or fine-grained, depending on the level of abstraction required for the analysis.  We present two kinds of control abstraction: (i) based on the source code structure of large software; and (ii) based on the structure of the finite state model. For the former, we propose the following levels of abstraction for control-oriented models: {\it package-level}, {\it class-level}, {\it method-level}, and {\it thread-level}.  For the latter, we present the concept of {\it path abstraction} as a way to construct reduced models by collapsing a sequence of transitions to a single transition.
	
	The concept of source-code control abstraction is fundamentally different from that of {\it inclusion filter} discussed in Section \ref{sec:JiveArch}.  The inclusion filter limits the full execution trace to a subtrace that is pertinent to the packages, classes and methods specified in the inclusion filter file.  On the other hand, the proposed abstractions for control-oriented models define reduced finite state models based on an analysis of the subtrace that is output by the inclusion filter. We elaborate on these abstractions below:
	
	\begin{algorithm}[H]
		\caption{Abstractions for Data-oriented Models}
		\label{alg:ModelAbstraction}
		\begin{algorithmic}  [1]
		\State \textbf{Input:} Execution trace $T[1..n]$, Key attributes \{$x_1,\cdots,x_m$\}
		\State \qquad\qquad Abstraction functions $f_1\dots f_m$
		\State \textbf{Output:} {Abstract State Model $G(V^a, E^a)$}
			\State $s_0$ = $\langle x_1=?,\cdots,x_m=? \rangle$;
			\State $V^a$ $\gets$  [$s_0$];  $E^a \leftarrow \{\}$;
			\State $s$ $\gets$ $s_0$;
			\For {\textbf{each} {\textit {FieldWrite Event} $x \leftarrow w$ in $T$ }}
			\State {\bf if} { $x$ $\not\in$ \{$x_1,\cdots,x_m$\} } {\bf then}
			\State \qquad {\bf continue};
			\State {\bf comment} Let $x$ be $x_i$ and {\it f}$_i$ be the abstraction function for $x_i$
			\State $s'$ $\gets$ $s$[$x_i \gets {\it f}_i(w)$]; 
			\State $V^a$  $\gets V^a$  $\cup$ \{$s'$\};  
			\State $E^a$  $\gets E^a$ $\cup$ \{$\langle$$s$, $s'$$\rangle$\};   
			\State $s$ $\gets$ $s'$;
			\EndFor
			\\
			{\bf return} $G(V^a, E^a)$;
		\end{algorithmic}
	\end{algorithm}

	\begin{enumerate}
		\item {\it Package-level. }
		It is the most abstract form and depicts the control flow among the various packages during execution. 
		Packages can be nested and this structure is indicated by the usual dot notation, e.g., {\tt p1.p2.p3.p4.c}, where 
		{\tt c} is a class name and {\tt p2, p3}, and {\tt p4} are the nested packages.
		\item {\it Class-level. }
		This is a lower level of control abstraction compared with the package-level and helps understand the runtime flow of control among classes, which may be spread across multiple packages.
		\item {\it Method-level. }
		This is the most refined form of control abstraction and depicts the runtime flow of control across methods. This control abstraction is suited for intra-package and intra-class analysis. 
		\item {\it Thread-level.}
		 
		 Here, the individual threads in a concurrent application can be singled out for analysis. A thread may touch upon multiple packages, classes and methods during its execution.
		  
	\end{enumerate}
	
	\begin{figure}[H]
		\begin{center}
			\includegraphics[width=\textwidth, height=5in]{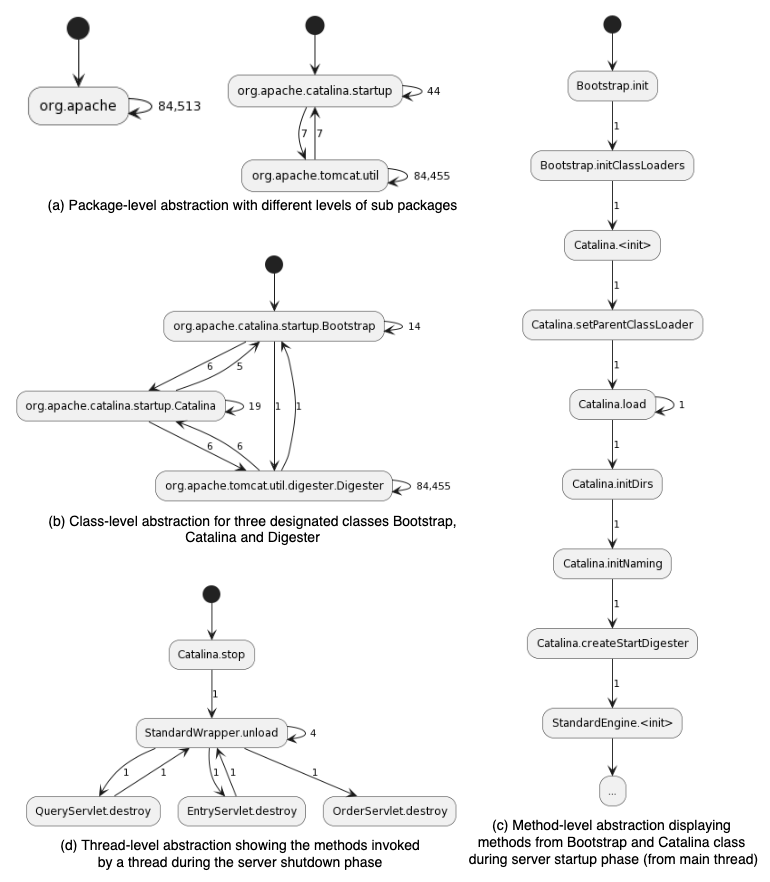}
			\caption{Sample abstractions for control-oriented models}
			\label{fig:control-abs}
		\end{center}
		\end{figure}
	  
        Figures \ref{fig:control-abs}a and \ref{fig:control-abs}b depict the different levels of package abstractions and class-level abstraction applied on the control-flow model of the execution of some of the classes in Apache Tomcat Server. Figures \ref{fig:control-abs}c and \ref{fig:control-abs}d show the method-level abstraction and thread-level abstraction during server startup and shutdown phase. This case study was discussed extensively in our earlier paper \cite{JJJS:FSM-SPE-2021} and section \ref{sec:casestudies} shows
        performance improvements due to model abstraction for this case study.  \\

   {\it Path Abstraction.}     
       In boolean-/multi-valued abstraction multiple states are collapsed to a single state, whereas in path abstraction multiple transitions are collapsed into a single transition. The concrete states are identified based on the path property and the abstract transitions between these states are added based on the paths between these states in the concrete model. The form of the path property to be verified determines the desired path abstraction.  
       
       Algorithm \ref{alg:PathAbstraction} shows how the abstract model is constructed based on path abstraction. Its inputs are an execution trace, the start state $s_0$ and boolean abstraction 
       functions $f_1$, $f_2$, and $f_3$ corresponding to a property {\tt P [$f_1\sim\sim>f_2\sim\sim>f_3$]}.  This property states that every path from a state where $f_1$ is true to a state where $f_3$ is true must pass through a state where $f_2$ is true.

     \begin{algorithm} [H]
		\caption{Path Abstraction for Control-Oriented Models}
		\label{alg:PathAbstraction}
         \textbf{Input:} Execution trace $T[1..n]$, Key (control) attribute \{$c$\}, start state $s_0$,\\ 
         and boolean abstraction functions $f_1,~f_2,~f_3$\\
		\textbf{Output:} {Abstract State Model $G(V^a, E^a)$}
		\begin{algorithmic}[1]
			\State $prev$ = $\langle c=?\rangle$;
			\State $V^a \leftarrow \{prev\};~~E^a \gets \{\};$
			\For {\textbf{each} {\textit {FieldWrite Event} $x \leftarrow v$ in T }}
			\State {\bf if} { $x$ $\neq$ $c$} {\bf then}
			\State \qquad {\bf continue};
			
            \State $curr$ $\gets$ $prev$[$c \gets v$];
            \State {\bf if} $f_1(curr) ~|| ~f_2(curr) ~|| ~f_3(curr)$ {\bf then} 
			\State \qquad $V^a$ $\gets$ $V^a$ $\cup$ $\{curr\}$; 
			\State \qquad $E^a$ $\gets$ $E^a$ $\cup$ $\{\langle prev,curr\rangle\}$; 
			\State \qquad $prev \gets curr$;
			\EndFor\\
			\Return $G(V^a, E^a)$;
		\end{algorithmic}
	\end{algorithm}
    \qquad 
       Algorithm \ref{alg:PathAbstraction} creates an abstract model with states $V^a$ and edges $E^a$ between these states. The abstract states are a subset of the concrete states satisfying the boolean functions ${\it f}_1~..~{\it f}_3$, respectively. For the path property to be meaningful, the boolean functions should partition the concrete states into three disjoint sets -- those that satisfy ${\it f}_1$,
       ${\it f}_2$ and  ${\it f}_3$ respectively.  
       (Note that the disjointness condition only applies for a specific {\tt P} property; different {\tt P} properties could have overlapping states with one another.)
       An edge $\langle s_i, s_j\rangle$ is present in $E^a$ if there is a path from concrete state $s_i$ to concrete state $s_j$.  If $s_1$, $s_2$, and $s_3$ are three states satisfying the abstraction functions ${\it f}_1$, {\it f}$_2$, and ${\it f}_3$, respectively,  a direct edge from $s_1$ to $s_3$  (i.e., without passing through $s_2$) would indicate a violation of the property {\tt P [$f_1\sim\sim>f_2\sim\sim> f_3$]}.

	\section{Property Specifications} \label{sec:lang}
	
	We now describe our proposed property specification language followed by examples of properties for several classical concurrency problems. The language can be used to formulate state-based and path-based properties over finite traces.  Although the language makes use of temporal operators, the standard CTL or LTL temporal logics \cite{Clarke:AutomaticVerification-FSCS-TLS-1986} are not applicable to our context because our execution traces are finite.  Even the language LTL$_{\it f}$ \cite{DV:LTLf-2013} for finite traces is not applicable because our properties are required to hold for a specific finite trace rather than over all possible finite traces.   
  Since our models are extracted from runs of a Java program, we provide a few syntactic conveniences in the property specification language -- operators for iteration, simple datatypes (integers, strings) and associated operators.
	
\subsection{Property Specification Language} \label{sec:psl}

We begin with a concise grammar below that summarizes the syntactic features in our property specification language, followed by a brief discussion of each feature and, finally, examples in section \ref{subsec:prop-examples}.

			\begin{tabular}{l c l}
				\it BoolExp  & ::=  &  {\it  BoolExp }  {\it boolop}  {\it BoolExp}  \\
				 &  &  $|$  {\it TemporalExp} $|$ {\it IterExp} $|$ {\it RelExp}  
                         $|$ {\tt (} {\it BoolExp} {\tt )} \\
                  & &    $|$ {\tt !} {\tt (} {\it BoolExp} {\tt )}  \\
                \it boolop & ::= & {\tt ->} $|$ {\tt \&\&} $|$ {\tt ||} \\
				\\ 
				\it  TemporalExp &  ::= &  {\tt F} { [{\it BoolExp }]} $|$ {\tt G}  [{\it BoolExp}] $|$  \\ 
                & & {\tt P} [ {\it BoolExp} $ \sim\sim>$   {\it BoolExp}  $\sim\sim>$   {\it BoolExp} ]  \\
		        \\
				\it IterExp    &  ::= &   {\tt all}(\it iter\_var, ListExp, BoolExp) $|$  \\	&& {\tt exists}\it  (iter\_var, ListExp, BoolExp) \\ \\
				\it  RelExp     &  ::= &  \it ArithExp relop ArithExp $|$ \it  StringExp sop StringExp $|$  \\ 
				        & & {\it ArithExp} {\tt in} {\it ListExp}  $|$ {\it StringExp} {\tt in} {\it ListExp} $|$ \\
                 && {\it iter\_var} {\tt in} {\it ListExp}  \\
                \it   relop   & ::= & = $|$ == $|$ != $| <  |  <= |  >  |  >=	$	\\
				\\
				\it  StringExp     &  ::= &    {\it string\_literal} $|$ \it str\_var  $|$ \it str\_var' \\
				\it  sop       &  ::= &   = $|$ == $|$ != \\
				\\
				\it  ListExp      &  ::=  &   {\it ArithExp:ArithExp} $|$ {\it list\_literal} $|$ \it list\_var $|$ list\_var' \\
				\it  listop   &  ::= &  {\tt min} $|$ {\tt max} $|$ {\tt size} \\
				\\
				\it   ArithExp &  ::= &  {\it  ArithExp} \it arithop {\it ArithExp } \\
				  &  &  $|$ \it  integer\_lit $|$ double\_lit $|$ ${\it arith\_var}$  $|$                             {\it  arith\_var' } \\ 
                    & &   $|$ {\it list\_var{\tt \#}listop} $|$ ( {\it ArithExp} ) \\
                    \it arithop  & ::= & {\tt +} $|$ {\tt -} $|$ {\tt *} $|$ {\tt /} \\\\
				
		\end{tabular}

	\begin{itemize}
		\item{\it Boolean Expressions.}
		 Boolean expressions are made of the logical connectives ({\tt ->, \&\&, ||, !})  which combine iterative, temporal and relational expressions to form complex properties.   
         Their order of precedence is {\tt !}, {\tt \&\&}, {\tt ||} and {\tt ->}, with {\tt !} highest and 
         {\tt ->} lowest.
		 
		\item {\it Iterative Expressions.} 
		Two overloaded operators, {\tt all} and {\tt exists}, are provided
		in order to state properties based on a range or a list of values.
		The iterator variable {\it iter\_var} takes on values from the range or list
		specified.
		\begin{enumerate}
			\item The {\tt all} operator checks whether {\it every} value in a numeric range or an explicit list satisfies
			a stated property--the third argument of {\it all}.
			 
			\item The {\tt exists} operator is similar to {\tt all} and checks whether {\it some}
			value in a numeric range or an explicit list satisfies a stated property--the third argument of {\it exists}.
		\end{enumerate}   

		\item {\it Temporal Expressions.}
		Three temporal operators are provided, two for stating state-based properties ({\tt F} and {\tt G}) and one for path-based properties ({\tt P}): 
		 \begin{enumerate}
   		  \item  {\tt F}[ {\it f} ] is true if {\it f} is true at some future state starting from the current state to the last state of the trace. 
        
		  \item  {\tt G}[ {\it f} ]  is true if {\it f} is true at every state from the current state to the last state of the trace.

         \item   {\tt P} $[ {\it f}_1 \sim\sim> {\it f}_2 \sim\sim> {\it f}_3 ]$  is true if {\it all} paths from a state where ${\it f}_1$ is true to a state where ${\it f}_3$ is true pass through a state where {\it f}$_2$ is true. 
         
		\end{enumerate} 
		Of the two state-based temporal operators, the {\tt G} operator is more common, as {\it invariance} ({\tt G}) is a more common property than {\it eventuality} ({\tt F}).  This is also the case for design-time model-checking, but it holds more strongly for runtime  verification because eventuality cannot be reliably checked with finite traces.

         Control-oriented models typically require properties to hold on different paths of the execution.  Hence we introduce the {\tt P} temporal operator for this purpose.  Note that the {\tt P} property is considered true if {\it f}$_1$ is false at the state at which the property is evaluated.  
         As the {\tt P} operator involves a universal quantification over paths, it cannot be expressed in terms of {\tt F} and {\tt G} which are fundamentally about states.
         
         A modified version of the {\tt P} operator can be used for quantification over \textit{some} paths, but it is less common in practice and hence we do not
         include it. 
         
     	\item {\it Next-State Variable.}  
	Another state-oriented temporal operator  is $x$', where $x$ is a key attribute of type arithmetic, string, or list and $x$' refers to its value in the next state (except at the last state).  The need to refer to the next state arises naturally in a state model built from a linear execution trace.  For distinct state models, there could be multiple next states for a state and therefore the property containing the expression $x$' should be true regardless of which next state is chosen to get the value of $x$. Unlike LTL's {\tt X[p]} operator, which can state a general property {\tt p} about the next state, the $x$' operator is applicable only to a variable, $x$, and allows one to state properties about the next state as well as properties that relate the current state and the next state, e.g., $x$ = $x$'.

	\item {\it Arithmetic and Relational Expressions.} 
	The familiar arithmetic and relational operators on integers and strings are supported.  A list can be an explicit literal or a numeric range of numbers.  List membership ({\tt in}) and simple numeric functions on lists ({\tt min}, {\tt max}, {\tt size}) are also supported.
	\end{itemize}
	
	\subsection{Property Specification Examples}
	\label{subsec:prop-examples}
	We present small examples drawn from classical problems in the literature.  The source codes and sample
	runs, along with outputs, for all examples are accessible at
	{\tt https://cse.buffalo.edu/jive/JSS\_Paper/}.
        We will return to these examples in subsequent sections when we discuss property verification on abstract models.

	\begin{enumerate}
		\item \textit{Dining Philosophers} \cite{Dijkstra:DP-1971}.
		The problem can be modeled by five threads where each thread represents one philosopher who repetitively goes through three states: {\tt T} (thinking), {\tt H} (hungry), and {\tt E} (eating). Assuming that the key attributes {\tt p1} .. {\tt p5} each hold one of these three values, the basic safety property -- that no two adjacent philosophers are eating concurrently -- can be stated as follows:\\
			
			{\tt G [ (p1 == "E" -> p2 != "E") \&\& \\
	\hphantom{ G }  (p2 == "E" -> p3 != "E") \&\& \\
	\hphantom{ G } 	(p3 == "E" -> p4 != "E") \&\& \\
	\hphantom{ G } 	(p4 == "E" -> p5 != "E") \&\& \\
	\hphantom{ G } 	(p5 == "E" -> p1 != "E") ] }\\
        
		\item \textit{Readers Writers} \cite{CHP:RW-1971}.	 Suppose a data resource is to be accessed by a number of reader threads and writer threads. Reader threads may access the resource concurrently, but a writer thread must access the resource exclusively.  Let the resource have two key attributes, {\tt r} and {\tt w}, which keep track of the number of active readers and writers respectively. Then the basic synchronization property can be stated as follows:\\
		
	    	{\tt G [ (r > 0 -> w == 0) \&\& \\
	\hphantom{ G }  (r >= 0) \&\& \\
	\hphantom{ G } 	(w == 0 || w == 1) ] }\\
	    \\
	    Priority for writer threads can be stated using an extra key attribute, {\tt ww}, which keeps track of the number of waiting writer threads.  Priority means that the number of active readers must monotonically decrease when there is a waiting writer. (Preemption of reader threads is not allowed.)   This requirement can be concisely stated using the next-state variable as follows:\\
	    
	    {\tt G [ (ww > 0 -> r' <= r)  ] }\\
		
       \item \textit{Elevator Problem.}
		Another classic example, the elevator is modeled using four key attributes: the current floor, {\tt f}, the direction of movement, {\tt d}, and two lists of numbers, {\tt up} and {\tt down}, which keep track of the outstanding requests from different floors.  The elevator operates in a reactive manner, repeatedly serving  requests on the {\tt up} and {\tt down} lists.  When all requests are served on both lists, the elevator comes to a stop at the last floor served. The elevator resumes motion once new requests arrive on either of these two lists.  Furthermore, the basic policy is that the elevator will serve all requests in one direction of movement before changing its direction of movement. These properties can be stated as follows:\\
		\begin{enumerate}
			\item Every request is eventually serviced.\\
			\\
			{\tt G [all(i, up, F[f == i]) ] \&\& \\
			G [all(i, down, F[f == i]) ]}\\
		
			\item If the elevator is moving down, it moves to lower numbered floors, and if the elevator is moving up, it moves to higher numbered floors. If there is a change in direction, then the elevator remains in the same floor.  \\
			\\
			{\tt G [ up==up' \&\& down==down' -> \\
			\hphantom{[ up==up'}	(d == "down"  \&\&  d' == "down" $\rightarrow$  f > f')  \&\&\\
			\hphantom{[ up==up'}	(d == "up"  \&\&  d' == "up" $\rightarrow$  f < f')  \&\&\\
			\hphantom{[ up==up'} (d != d' $\rightarrow$  f == f') ]}\\

   In the above specification, since only one state component may change at a time, the check {\tt up==up' \&\& down==down'}  ensures that the conditions on {\tt d} and {\tt f} are correctly checked, especially because they refer to the values of {\tt d} and {\tt f} in the next state. \\
   
			\item The elevator changes direction only when there are no more requests in the direction of movement on both lists.\\\\
			{\tt G [ up == up' \&\& down == down' ->\\
			\hphantom{up == up'} (d == "down"  \&\&  f <= up\#min \&\& f <= down\#min\\
			\hphantom{up == up' down == down'} ->  d' == "up") ]}\\
			{\tt G [ up == up' \&\& down == down' ->\\
			\hphantom{up == up'} (d == "up" \&\& 	f >= up\#max \&\&	f >= down\#max \\
			\hphantom{up == up' down == down'} -> d' == "down")	]}\\
		\end{enumerate}
  
		\item \textit{Path Properties.}  Path properties are examples of control-oriented properties and they arise in a natural way in different application contexts.  The most common is the authorization context, where we want to state the condition that every request to use a certain resource must be authorized before the resource can be used.  Request, authorization, and usage are performed in different modules and we want the overall flow of control to satisfy the following property, where the key attribute {\tt s} gives the control location. \\\\
		
	    {\tt P [s == "Request" $\sim\sim$> s == "Authorise" $\sim\sim$> s == "Use"]} \\

        We discuss this property further in Section \ref{sec:casestudies} under the OAuth Protocol case study.\\
        
        For a different application scenario, in a data analysis application, data must  be cleaned after it is acquired and before it can be analysed. This requirement can be stated by a {\tt P} property that mandates every path from data acquisition to data analysis to pass through the data cleaning process, as follows:\\\\
		 
	    {\tt P [s == "Acquire" $\sim\sim$> s == "Clean" $\sim\sim$> s == "Analyse"]}\\
     
    HTTP request processing using servlets also provides a rich setting for path properties, as discussed in our previous work \cite{JJJS:FSM-SPE-2021}.  For example, a common requirement is that once an object is initialized, it must process requests before it is destroyed.  That is, every path from initialization to destruction must pass through request processing. \\

	    {\tt P [s == "Init" $\sim\sim$> s == "Serve" $\sim\sim$> s == "Destroy"]} \\
   
	\end{enumerate}
	
	\section{Property Checking on Abstract Models}
	\label{sec:verification}

	We begin with a discussion on linear and distinct state models. The linear state model is the most accurate model for an execution trace and facilitates property checking for all forms of expressions within the limits of a finite trace.  
	The distinct state model is constructed by merging duplicate states in the linear state model.  This reduces the state space and thereby enhances the efficiency of checking {\tt G} properties. 	 
  
        Property checking on the linear state model is straightforward. The core of the technique is a set of cases for the temporal 
        operators {\tt G}, {\tt F}, and {\tt P}.  For {\tt G[p]} to be true at a state $s$, the expression {\tt p} must be true at $s$ and every subsequent state. For {\tt F[p]} to be true at a state $s$, the expression {\tt p} must be true at $s$ or some subsequent state.
        For {\tt P[p$\sim\sim>$ q$\sim\sim>$ r]} to be true at a state $s$, if {\tt p} is true at $s$ and {\tt r} is true at some subsequent state $t$, then {\tt q} must be true at some state between $s$ and $t$.
        
		The evaluation of a propositional expression in some state makes use of the values of key attributes in that state along with a simple case analysis on the types of operators (logical, arithmetic or relational) for each subexpression. The datatypes supported are integer, double and string, and the operators are overloaded to perform operations based on the datatype of the operands. 
		
		 The iterative operators, {\tt all}({\it iter\_var, ListExp, Exp}) and {\tt exists}({\it iter\_var, ListExp, Exp}), are also evaluated using the values of key attributes in a given state.  They check whether the given expression {\it Exp} is true at the current state for every/some value in the sequence of values specified by {\it ListExp}.  The list expression can be a numeric range or an explicitly defined list.  The control variable for the iteration is {\it iter\_var}, which repeatedly is bound to the values in the given range or list.
		
		In the remainder of this section we focus on property checking on abstract
		models. As described in section \ref{sec:abstraction}, for data-oriented properties ({\tt G} properties), abstract states are generated based on boolean or multi-valued abstractions, and transitions connecting these abstract states are created. For control oriented properties ({\tt P} properties), concrete states were first identified (from the {\tt P} property) and abstract transitions between these states are created. 
		
	We note at the outset that not all forms of properties or abstractions are compatible with property checking on abstract models.
	
	\begin{itemize}
	    \item The merging of states during abstraction often leads to an overestimation of paths between states and hence the abstract state models are not suitable for checking {\tt F} and {\tt P} properties. 
	    \item The expression $x$' (the value of $x$ in the next state) cannot be correctly interpreted in abstract state models because non-adjacent states of the linear state model can become adjacent in the abstract state model.  
	\end{itemize}
	
	For these reasons, the data-oriented properties considered are of the form {\tt G [$p$]}, where property $p$ is a propositional formula possibly augmented with numeric and string datatypes and associated operators for which validity can be decided.  We also consider path properties of the form {\tt P[}$p_1\sim\sim>p_2\sim\sim>p_3${\tt ]}. 

	\subsection{Property Checking for Boolean Abstractions}\label{sec:verification-asm}
  
  We first present property checking for boolean abstractions and then explain how this approach can be extended to multi-valued abstraction.  An abstract state model is obtained by mapping a concrete state of values $\langle v_1,...,v_n\rangle$ to an abstract state  $\langle f_1(v_1),....,f_n(v_n)\rangle$, where $f_1,..,f_n$ are abstraction functions.   
  Not all boolean abstractions are compatible with property checking.  For example, if an integer-valued key attribute, $k$, was abstracted as two values, negatives ($k<0$) and non-negatives ($k\ge0$), a property that seeks to check for a more specific value for $k$, e.g., $k = 1$, cannot be validated using the abstract model.  In practice, we examine the property to be checked and devise an abstraction that is compatible with the property. Sometimes, it might be just easier to work with the concrete model and bypass abstraction altogether.

\begin{definition}
    Given a set of key attributes $v_1...v_n$ and boolean abstraction functions $f_1 ... f_n$.
     If {\tt G[p]} is to be verified on a concrete state model, then
     the property to be verified on each state of the abstract state model is:
 
      $$g_1(v_1) \land  ... \land g_n(v_n) \rightarrow {\tt p}$$ 
     
   \noindent where each $g_i(v_i)$ is defined as $f_i(v_i)$ or $\neg f_i(v_i)$, depending upon 
   whether the value of the $i$-th state component in the abstract model is {\it true} or {\it false}.
\end{definition}
   
  The boolean abstraction functions $f_i (v_i)$ are simple relational expressions such as $v_i=c$, $v_i \neq c$, $v_i<c$ or $v_i>c$ or a composition of these simple relational expressions and boolean connectives ($\land$, $\lor$, $\neg$). Algorithm \ref{alg:ASM-PC} represents the property checking on abstract state model.
   
   \begin{algorithm} [H]
 	\caption{Property Checking with Boolean Abstraction Functions }
 	\label{alg:ASM-PC}
 	\textbf{Input:} Abstract State Model $G(V^a, E^a)$ , Property {\tt G[p]}, \\
\hphantom{ GGG  } 	Key Attributes $\langle v_1,...,v_n\rangle$, Boolean Abstraction Functions $f_1,...,f_n$.	\\
 	\textbf{Output:} boolean  (verified/not verified)
	\begin{algorithmic}[1]
		\State Let each abstract state $s \in V^a$ be $\langle b_1,...,b_n\rangle$, where $b_i$ is a boolean.
		\State {\bf boolean} $output$ = {\tt false};
		\State  {\bf for each}  $s \in V^a$
		\State \qquad {\bf for each} $b_i \in s$ 
		\State  \qquad \qquad  \qquad {\bf if} ($b_i$) {\bf then}  $g_i$ $\leftarrow$ $\llbracket f_i(v_i) \rrbracket$ {\bf else} $g_i$ $\leftarrow$  $\llbracket \neg f_i(v_i) \rrbracket$
		\State \qquad {\bf end for}
		\State \qquad {\bf if} $\neg$ \Call{Valid}{$\llbracket g_1\land...\land g_n \rightarrow {\tt p} \rrbracket$} {\bf then}
				\State \qquad \qquad \Return {\tt false}; \qquad {\bf comment} Property {\tt p} failed at state $s$
		\State {\bf end for}
		\State \Return {\tt true}; \qquad\qquad\qquad {\bf comment} Property {\tt p} is valid for all states
	\end{algorithmic} 
\end{algorithm}

       \begin{figure}[h]
   	\begin{center}
   		\includegraphics[width=\textwidth]{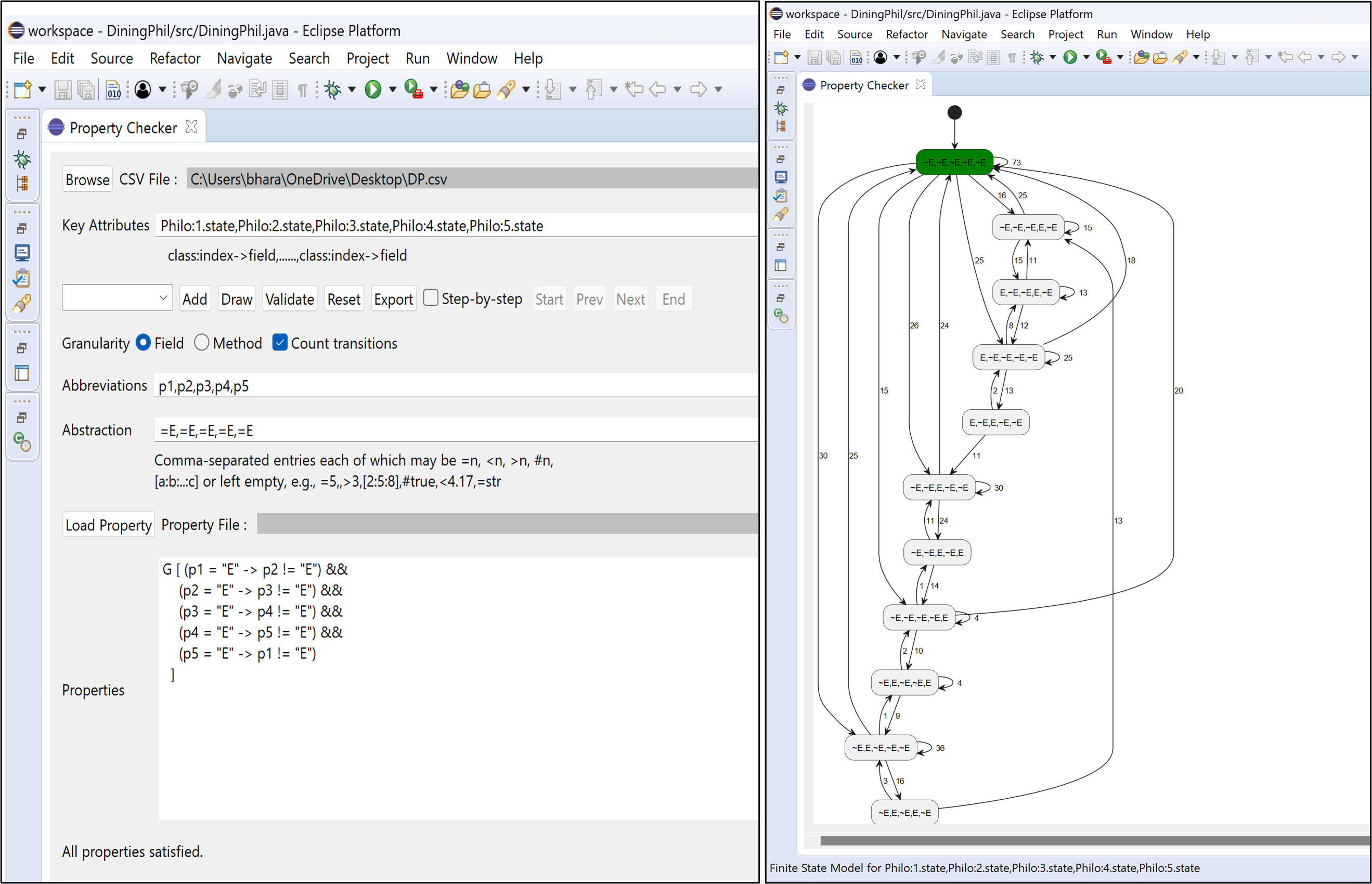}
   		\caption{Screenshot of JIV$^2$E for off-line property-checking on a runtime abstract model of the Dining Philosophers program execution. The boolean abstraction function is: {\tt p}$_i$=={\tt E}, for $i=1..5$. All 11 abstract states are shown in the figure.  Each state component is abbreviated by the value 
   		{\tt E} or $\sim${\tt E}, where the latter abstracts the concrete values {\tt T} and {\tt H}. The green-highlighted state is the start state. }
   		\label{fig:abs-DP}
   	\end{center}
   \end{figure}

\paragraph {Dining Philosophers Problem (section \ref{sec:lang})}  The safety requirement shown
below is amenable to a boolean abstraction:  {\tt p}$_i$ {\tt == "E"}, for $i=1..5$. 
Figure \ref{fig:abs-DP} shows a screen-shot of off-line property-checking with abstraction in JIV$^2$E.  (Later, in Figure \ref{fig:jivearch} we illustrate how online property-checking with abstraction facilitates early error detection for an erroneous coding of this example.)  The components of the abstract states are shown as {\tt E} or {\tt $\sim$E}, where {\tt $\sim$E} abstracts {\tt T} and {\tt H}.
\\
\\
			{\tt G [ (p1 == "E" -> p2 != "E") \&\& \\
	\hphantom{ G }  (p2 == "E" -> p3 != "E") \&\& \\
	\hphantom{ G } 	(p3 == "E" -> p4 != "E") \&\& \\
	\hphantom{ G } 	(p4 == "E" -> p5 != "E") \&\& \\
	\hphantom{ G } 	(p5 == "E" -> p1 != "E") ] }\\

\noindent
The maximum number of abstract states (in a correct implementation) will be just 11--five abstract states where exactly one philosopher is eating; five abstract states where exactly two philosophers are eating; and one abstract state where no philosopher is eating. For each state, depending upon the abstract values in that state, a specific property needs to be constructed, as shown
in Algorithm \ref{alg:ASM-PC}, and this property needs to be validated.  We illustrate the property checking for the abstract state {\tt (E,$\sim$E,$\sim$E,E,$\sim$E)}. The overall property to be validated for this state is: \\
  
\noindent {\tt  (p1 = "E")} {\tt \&\&}  {\tt (p2 != "E")} {\tt \&\&}  {\tt (p3 != "E")} {\tt \&\&}
          {\tt  (p4 = "E")} {\tt \&\&} {\tt (p5 != "E")}\\
\hphantom{ G } $\rightarrow$ \\
	\hphantom{ G G }	{\tt  (p1 == "E" -> p2 != "E") \&\& \\
	\hphantom{ G G }  (p2 == "E" -> p3 != "E") \&\& \\
	\hphantom{ G G } 	(p3 == "E" -> p4 != "E") \&\& \\
	\hphantom{ G G } 	(p4 == "E" -> p5 != "E") \&\& \\
	\hphantom{ G G } 	(p5 == "E" -> p1 != "E")  }\\

\noindent
In a similar manner, for each of the other 10 abstract states, an appropriate property is constructed (by Algorithm \ref{alg:ASM-PC}) and used in place of the antecedent of the above property (i.e., to the left of top-level $\rightarrow$).  It is easy to see that these properties are all valid and thus the given {\tt G} property is valid for all concrete states of a run of (a correct implementation of) the Dining Philosophers problem.

 \paragraph{Readers-Writers Problem (section \ref{sec:lang})} 
	This example illustrates the use of multi-valued abstraction and it also shows how through a simple rewriting of the {\tt G} property, it is possible to use boolean-valued abstraction.  The concurrency requirement stated below is not immediately amenable to boolean abstraction because we need to distinguish three ranges of integers for {\tt r} (negatives, zero, and positives)
    due to the conditions {\tt r > 0} and {\tt r >= 0}.  We show first how the property can be rewritten so that boolean-valued abstraction can be used and, subsequently, we show how the property can be checked with multi-valued abstraction.\\
  
	 {\tt
		G [ (r > 0 -> w == 0) \&\&\\ 
\hphantom{ G G }		(r >= 0) \&\&\\
\hphantom{ G G }		(w == 0 || w == 1)
		]
	} \\
	 
\noindent	
Because the {\tt G} operator distributes over {\tt \&\&} we can rewrite the property as follows. (Table
\ref{tab:properties equiv} shows the more commonly needed properties.)
		 
{\tt
		G [ r > 0 -> w == 0 ] \&\&\\
\hphantom{ G}		G [ r >= 0 ] \&\&\\
\hphantom{ G}		G [ w == 0 || w == 1  ] } \\
	
	The individual {\tt G} properties are now amenable to boolean abstractions:\\ 
	{\tt G [ (r > 0 -> w == 0)]} is amenable to the abstraction {\tt r > 0} and {\tt w == 0};  \\
	{\tt G [ r >= 0 ]} is amenable to {\tt r >= 0}; and \\ 
	{\tt G [ w == 0 || w == 1 ]} is amenable to {\tt w == 0 || w == 1}.  \\
	Using this approach, three different abstract models need to be constructed but property-checking is done with a smaller property. Next we will discuss how property checking can be done with multi-valued abstractions.

\begin{figure}[tbh]
		\begin{center}
			\includegraphics[width=\textwidth, height = 4.5in]{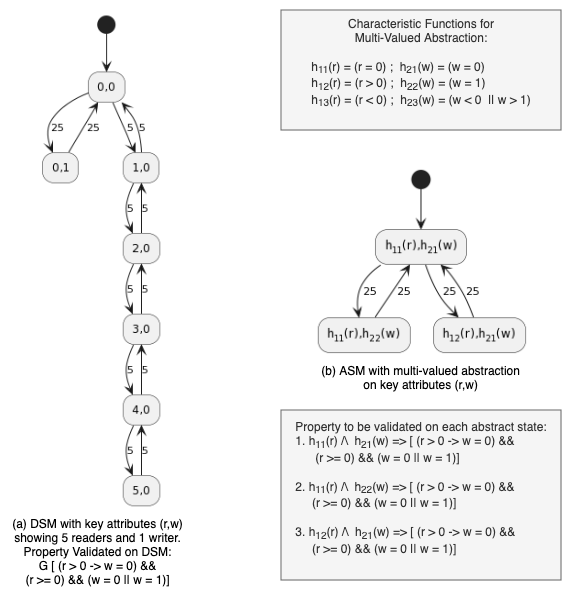}
			\caption{Property Checking on Abstract State Models with Multi-valued Abstractions}
			\label{fig:ASM-RW-Abs}
		\end{center}
	\end{figure}

		\begin{table}[tbh]
		\caption{Equivalence of Properties }
		\label{tab:properties equiv}
		\begin{center}
		\begin{tabular}{l c l}
			\toprule[1pt]
				G[$P_1\land ...\land P_n$]  & $\equiv$ & G[$P_1$]$\land...\land$ G[$P_n$]   \\
				G[$P \rightarrow (Q_1\land ...\land Q_n$)]  & $\equiv$ & G[$P\rightarrow Q_1$]$\land...\land$ G[$P \rightarrow Q_n$]   \\
				G[$P_1\lor ...\lor P_n$]  & $\not\equiv$ & G[$P_1$]$\lor...\lor$ G[$P_n$]   \\

				\midrule	
				P[($P_1 \lor P_2$) $\sim\sim>P_3 \sim\sim> P_4$] & $\equiv$& P[$P_1\sim\sim>P_3 \sim\sim> P_4$] $\land$  \\ & &  P[$P_2\sim\sim>P_3 \sim\sim> P_4$]\\
				
				P[$P_1\sim\sim>( P_2 \lor P_3) \sim\sim> P_4$] & $\not\equiv$& P[$P_1\sim\sim>P_2 \sim\sim> P_4$] $\lor$  \\ & &  P[$P_1\sim\sim>P_3 \sim\sim> P_4$]\\
				
				P[$P_1\sim\sim>P_2 \sim\sim> (P_3 \lor P_4)$] & $\equiv$& P[$P_1\sim\sim>P_2 \sim\sim> P_3$] $\land$  \\ & &  P[$P_1\sim\sim>P_2 \sim\sim> P_4$]\\
			   \bottomrule	
			 
			\end{tabular}
		\end{center}	
	\end{table}

\subsection{Property Checking for Multi-valued Abstractions}
\label{sec:verification-asm-mva}

The key idea underlying property checking with multi-valued abstraction functions is to define a {\it characteristic function} for each abstract value.  Consider a key attribute $v$ and its concrete domain $C$.  The abstraction function effectively partitions $C$ into a set of disjoint sets, where each disjoint set can be thought of as representing one abstract value. The characteristic function for a particular abstract value is a boolean-valued function, $C\rightarrow B$, which returns true for each concrete value that corresponds to the abstract value, and false otherwise.  We elaborate on this idea below.

\begin{definition}
    Given a key attribute $v$ and an abstraction function that partitions its
     concrete domain $C$ into $j$ disjoint sets. We introduce a domain of abstract values 
    $D = \{d_1,...,d_j\}$, where each abstract value is associated with one distinct disjoint set.
    For each abstract value $d_i$,
    for $i=1..j$, we define a characteristic boolean-valued function $h_i:C\rightarrow B$, such that $h_i(x) = true$ if $x$ belongs to the concrete set abstracted by $d_i$; and   $h_i(x) = false$, otherwise.
    
    Given a set of key attributes $v_1...v_n$ with abstract attribute domains $D_1 = \{d_{11},...d_{1j_1}\}$, ..., $D_n=\{d_{n1},...d_{nj_n}\}$ respectively.  We define characteristic functions for the abstract values in each attribute domain in a similar manner.  Then, given a property {\tt G[p]} to be verified on the concrete model, the property to be verified on each state of the abstract model is:
 
      $$g_1(v_1) \land  ... \land g_n(v_n) \rightarrow {\tt p}$$ 
     
   \noindent where each $g_i(v_i)$ is defined as one of $h_{i1}(v_i)$ ... $h_{ij_i}(v_i)$, 
   according to whether the abstract value for $v_i$ was one of $d_{i1} .,. d_{ij_i}$. 
\end{definition}

As an example of the characteristic functions, consider the multi-valued abstraction (1:3:5) shown in
Figure \ref{fig:lsm-dsm-asm}(d).  The four abstract values for the key attribute {\tt k} can be characterized by four boolean-valued functions, as follows:\\

{\tt h$_1$(k) = k < 1;} \qquad\qquad\qquad\qquad\qquad {\tt \qquad h$_2$(k) = k >= 1 \&\& k < 3} \\
	\hphantom{ H }{\tt h$_3$(k) = k >= 3 \&\& k < 5} \qquad\qquad\qquad \hphantom{G  } {\tt h$_4$(k) = k >= 5}\\

In the light of the above definition, Algorithm \ref{alg:ASM-PC} can be generalized for multi-valued abstraction functions as shown in Algorithm \ref{alg:ASM-PC-MV}.  We revisit the Readers-Writers problem and show in Figure \ref{fig:ASM-RW-Abs} how the property for this problem can be checked with multi-valued abstractions.

	\begin{algorithm} [H]
 	\caption{Property Checking with Multi-valued Abstraction Functions }
 	\label{alg:ASM-PC-MV}
 	\textbf{Input:} Abstract Model $G(V^a, E^a)$ , Property {\tt G[p]},\\ 
\hphantom{ GGG  } 	Key Attributes: $\langle v_1,...,v_n\rangle$, \\
\hphantom{ GGG  } 	Abstract values: $D_1 = \{d_{11},d_{12},...\},...,D_n = \{d_{n1},d_{n2},...\}$ \\
\hphantom{ GGG  }   Boolean Abstraction Functions:   $\{h_{11},h_{12},...,h_{n1},h_{n2},...\}$, \\
 	\textbf{Output:} boolean (verified/not verified)
 	\begin{algorithmic}[1]
	     \State Let each abstract state $s \in V^a$ be $\langle a_1,...,a_n\rangle$, where $a_i \in D_i$, for $i=1..n$.
		\State {\bf boolean} $output$ = {\tt false};
		\State  {\bf for each} $s \in V^a$
		\State \qquad {\bf for each} $a_i \in s$ 
		\State  \qquad \qquad {\bf case} ($a_i$) {\bf of}
		\State \qquad\qquad\qquad $d_{i1}$ : $g_i$ $\leftarrow$ $\llbracket h_{i1}(v_i) \rrbracket$  
		 \State \qquad\qquad\qquad    ...
		  \State \qquad\qquad\qquad $d_{ij_i}$ : $g_i$ $\leftarrow$  $\llbracket h_{ij_i}(v_i) \rrbracket$
		\State \qquad \qquad {\bf end case}
		\State \qquad  {\bf end for}
		\State \qquad {\bf if} $\neg$ \Call{Valid}{$\llbracket g_1\land...\land g_n \rightarrow {\tt p} \rrbracket$} {\bf then}
				\State \qquad \qquad \Return {\tt false}; \qquad {\bf comment} Property {\tt p} failed at state {\tt s}
		\State {\bf end for}
		\State \Return {\tt true}; \qquad\qquad\qquad {\bf comment} Property {\tt p} is valid for all states
		
	\end{algorithmic} 
\end{algorithm}
	
\subsection{Property Preservation}  

We provide an informal explanation of the correctness of the abstraction algorithms presented in this section.  For checking a {\tt G[p]} property on the concrete model, we need to check for each concrete state:

({\it v}$_1$ = {\it c}$_1$ $\wedge$ ... $\wedge$ {\it v}$_n$ = {\it c}$_n$) $\rightarrow$ {\tt p}

\noindent 
where 
{\it v}$_1$ ... {\it v}$_n$ are the key attributes that constitute the state vector and
{\it c}$_1$ ... {\it c}$_n$ are their respective values for that concrete state. 
In going from the concrete to the abstract model, the form of the above implication remains the same, but each
conjunct is replaced by a more general condition because each abstract value stands for
a set of possible concrete values.

A boolean abstraction function, $f_i(v_i)$, partitions the $i^{th}$ concrete domain into two sets - those values that satisfy $f_i(v_i)$ and those that satisfy $\neg f_i(v_i)$.  For each abstract state of boolean values, a specific property for that state needs to be formulated and its validity checked.  If the $i^{th}$ component of an abstract state is {\it true} the condition $f_i(v_i)$ is used as the $i^{th}$ conjunct of the implication for this state; otherwise, $\neg f_i(v_i)$, is used.  If, for every state, the property so formulated is valid, then property {\tt p} is true for all concrete states and hence {\tt G[p]} is true.  This is precisely what Algorithm \ref{alg:ASM-PC} does.

Multi-valued abstraction functions are a generalization of boolean abstraction functions. Given a set of key attributes $v_1,...v_n$ with abstract domains $D_1,...,D_n$ respectively.   Each value of the abstract domain $D_i$ = $\{d_{i1} ... d_{ij_i}\}$ is associated with a characteristic function, $h_{i1} ... h_{ij_i}$, respectively.  That is, $h_{ik}(v_i)$, for $k=1..j_i$, is true when $v_i$ has value $d_{ik}$, and false otherwise.  Now, for each abstract state of values, a specific property for that state needs to be formulated and its validity checked.  If the abstract value for $v_i$ is $d_{ik}$, for some $k=1..j_i$, the condition $h_{ik}(v_i)$ is used as the $i^{th}$ conjunct of the implication for this state.  If, for every state, the property so formulated is valid, then property {\tt p} is true for all concrete states and hence {\tt G[p]} is true.
This is the basis for Algorithm \ref{alg:ASM-PC-MV}.

Property-checking on control-oriented models is considerably simplified thanks to construction of the abstract model in Section \ref{subsec:control-oriented}.  Every edge in the abstract model stands for a path in the concrete model.
For a path property of the form {\tt P [$f_1\sim\sim>f_2\sim\sim>f_3$]}, Algorithm 
\ref{alg:ASM-PC-Path} simply carries out a test for the presence of a direct edge from the abstract state
$s_1$ (for concrete states satisfying $f_1$) to the abstract state $s_3$ (for concrete states satisfying
$f_3$).  The presence of such a direct edge indicates that there is a property violation, i.e., there is a path from a concrete state satisfying $f_1$ to a concrete state satisfying $f_3$ without going through a concrete state satisfying $f_2$.

	\begin{algorithm} [H]
 	\caption{Property Checking for Path Abstraction }
 	\label{alg:ASM-PC-Path}
 	\textbf{Input:} Abstract Model $G(V^a, E^a)$ , Property {\tt P [$f_1\sim\sim>f_2\sim\sim>f_3$]}\\
 	\textbf{Output:} boolean
 	\begin{algorithmic}[1]
	    \State Let $V^a = \{s_0,s_1,s_2,s_3\}$;
	    \State {\bf comment} Abstract states $s_1..s_3$ satisfy functions $f_1..f_3$ respectively.
		\State {\bf if} $ \langle s_1,s_3 \rangle \in E^a$
		\State \qquad{\bf return} false;
		\State {\bf else return} true;
	\end{algorithmic} 
\end{algorithm}

\section{Handling Large Scale Software} \label{sec:casestudies}

We highlighted earlier in Section \ref{sec:abstraction} two broad classes of finite state models: data-oriented models and control-oriented models.  Data-oriented models are primarily used for intra-class analysis, i.e., the analysis of state information within classes at lower levels of the software.

 Control-oriented models are useful in understanding the flow of control among modules at 
 higher-levels of the software.  These levels are typically
  developed with {\it data abstraction} as a key principle.  That is, data representations are hidden in the lower levels of design and the higher levels expose only an interface of methods. Thus, control-oriented models are useful in unraveling the working of a large software.
  We perform this analysis  starting from the main class/method as given in the {\it manifest file} of the application.

  As noted in the introduction, the applications that are pertinent to our study are concurrent systems such as servers and controllers that operate in a cyclic manner and are amenable to finite-state analysis.  We therefore choose one application of a data-oriented model (Multirotor Drone Controller) and one application of a control-oriented model (Open Authorization Protocol). We first present an overview of online verification architecture, followed by a discussion of these two application case-studies.

\subsection{Online Verification Architecture}
\label{sec:JiveArch}
\begin{figure}[tbh]
		\includegraphics[width=\textwidth]{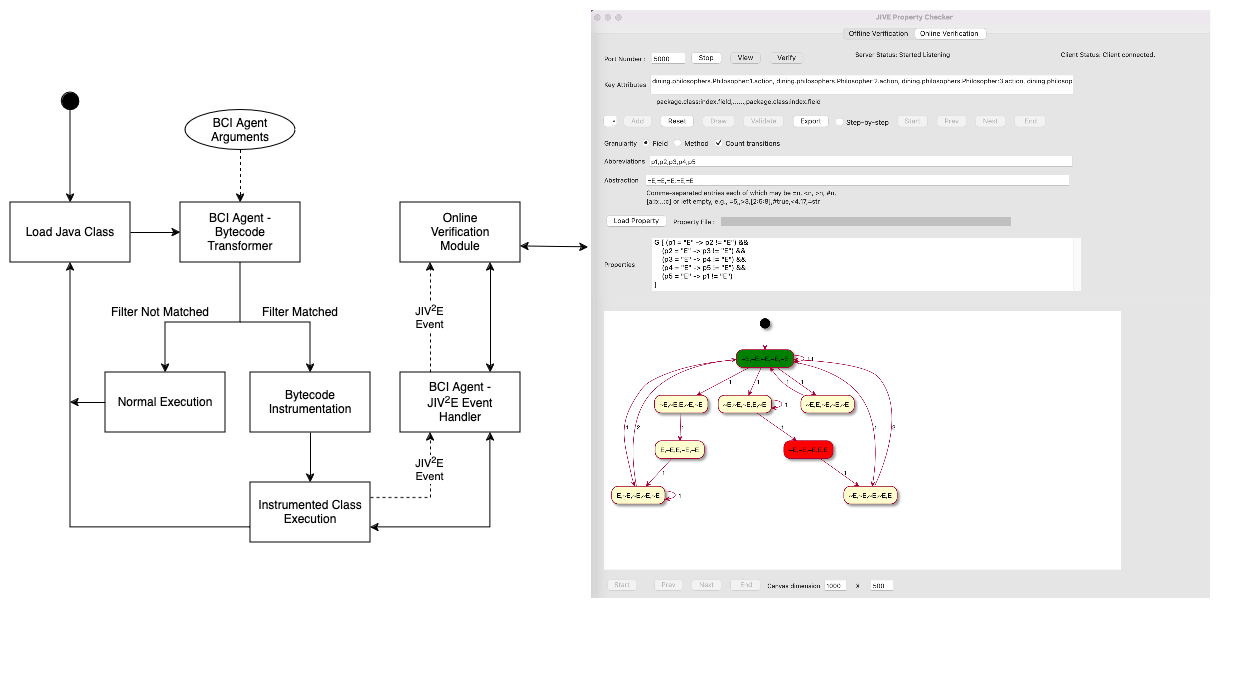}
		\caption{The JIV$^2$E Online Verification Architecture with Byte-Code Instrumentation. The User Interface shows the abstract state diagram for an erroneous run of the Dining Philosophers problem. The property checker highlights in red the erroneous state as it occurred.}
		\label{fig:jivearch}
	\end{figure}

	A key component of the online verification architecture is the efficient generation of events as this is crucial for the real-time construction of finite state models for long executions.  Towards this end,
    the overall architecture of JIV$^2$E incorporates byte-code instrumentation and
    inclusion filters for efficient event generation (see Figure \ref{fig:jivearch}).  
    This approach enables generation of events orders of magnitude faster than is possible with the standard JPDA.
    Essentially, using the Java instrumentation API \cite{javalanginstrumentJavaSE8} and ASM library \cite{ASM2002},\cite{ASM2007}
    the bytecodes of Java classes are modified (without changing their source codes) during the loading of classes, i.e., prior to execution.  At class-load time, additional instructions for generating JIV$^2$E events are injected into those classes that have been chosen for instrumentation.
	
	  The bytecode instrumentation code is deployed as a Java agent, called \textit{BCI agent}, which is first loaded into the JVM during the application startup. The {\it BCI agent}  has two components: {\it bytecode transformer} and the JIV$^2$E {\it  event handler}. The BCI agent's bytecode transformer module is responsible to instrument the loaded classes matching the 
	 {\it inclusion filter}; and the JIV$^2$E event handler module is responsible to collect the events emitted by the instrumented classes and send them to the online verification module for processing. It also responds appropriately to the messages from online verification module.

	The JVM invokes the BCI Agent's bytecode transformer through its {\it premain} method, which is similar to the {\it main} method in regular Java programs. The {\it premain} method receives two arguments: a  string object {\it agent arguments} which contains the user arguments to the agent; and an instance of the {\it instrumentation} class \cite{javalanginstrumentJavaSE8}. 
	The agent arguments are processed by the BCI Agent in order to obtain various inputs required for instrumentation. The agent is initialized with these argument values or with the default values as required. The mandatory argument is the inclusion filter file, which enables selective instrumentation of the application.
    The optional arguments are the execution trace file, online verification module's IP address and port number, enable/disable logging flag and event buffer size using which the BCI Agent can be customized. 
	
	User-defined bytecodes can be injected into each class that is loaded using a {\it transformer} method in the {\it instrumentation} class.  Essentially, this method weaves the required bytecode instructions so that pertinent context information (current thread name, class name, line number, etc.) can be obtained in order to construct the JIV$^2$E event. For online verification, the JIV$^2$E event handler delivers the events generated by the instrumented classes to the verification module. This event handler can optionally write the events to a trace file for offline analysis. The events can also be optionally buffered depending on the event buffer size, before sending the events to the verification module.
	
	The online verification module receives the events as the instrumented classes execute and performs model construction and property checking.  Only events pertaining to user-defined {\it key attributes} are used for model construction and abstraction can be performed as events are received. The abstraction criteria are based on the properties to be verified and are defined on a per attribute basis.  Online verification is primarily applicable to state-based properties (stated using the {\tt G} operator) because path-based properties can refer to future states that are not yet encountered.
	
    The online verification module communicates the property violation to the JIV$^2$E event handler module, which can optionally terminate the application depending on the severity of the property violated.
 	Although BCI is a well-known technique in the literature, our contribution lies in the use of this technique for customized generation of JIV$^2$E events and achieving a scalable efficient finite state model construction.

\subsection{Multirotor Drone Controller}
JMAVSim \cite{JMAVSim1, JMAVSim2} is an open-source multirotor drone controller simulator that allows users to test and fly copter-type vehicles based on PX4 \cite{PX4conf,PX4}. Developed in Java, it is primarily used for testing various actions like takeoff, flight, and landing of the vehicle. It can also simulate different failure conditions such as GPS failure and battery drain.  

In this study, we use JMAVSim to simulate the flight of an unmanned aerial vehicle (UAV) over a circular path, starting from the center of the circular path as its home location, denoted by point (C). The UAV takes off from its home location (C) at an altitude range of 322-324m and ascends to a height of about 40m before flying towards point (P) on the circular path, at an altitude range of 362-365m. It then rises another 20m to reach an altitude range of 385-386m and starts moving along the circular path. After completing one lap, the UAV climbs an additional 25m to reach an altitude range of 410-411m and performs a second round. Upon completion of the second round, the UAV descends to an altitude range of 362-364m and returns to the center of the circle (C). 

A data-oriented model for the UAV can be constructed which captures the GPS coordinates during its flight. The \textit{key attributes} selected to construct the model are: \textit{latitude}, \textit{longitude} and \textit{altitude}. The flights at various altitudes are obtained by applying range abstraction, as the UAV's movement is always in an altitude range rather than at a fixed altitude value. Additionally, to convert the latitude and longitude coordinates into compass directions, a custom abstraction function is employed, which maps the coordinates to their respective directions relative to the starting point.

\begin{figure}
         \centering
         \includegraphics[height=7.5in]{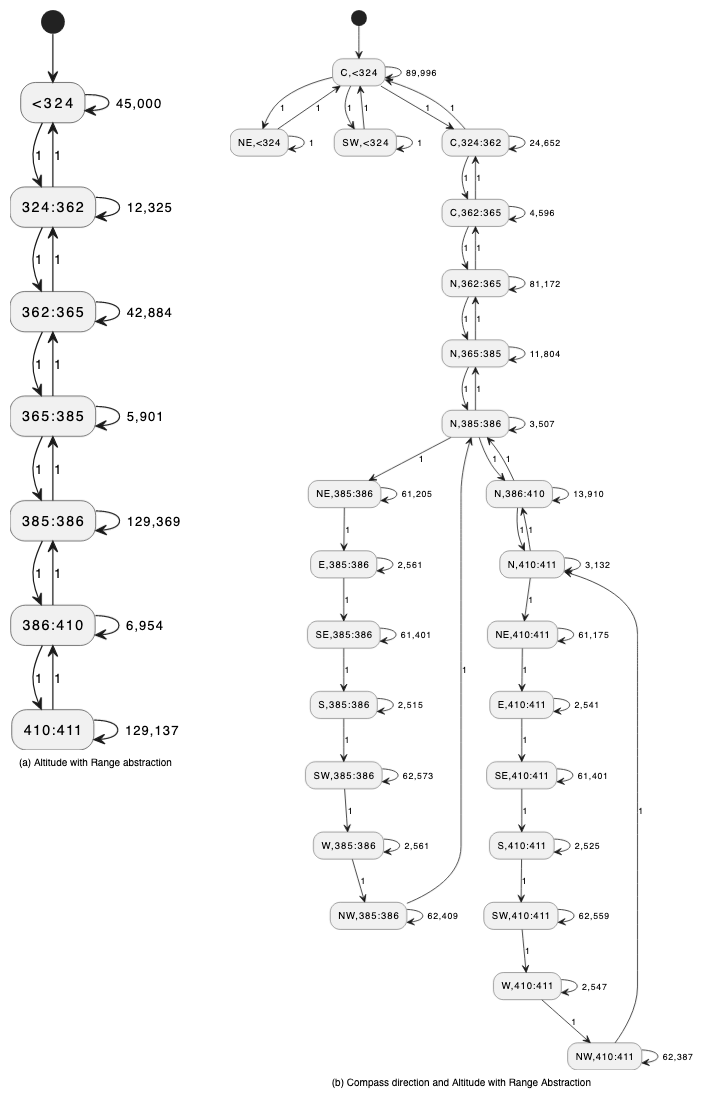}
         \caption{Combined Abstract model depicting the  navigation of the UAV in a circular path at different altitudes, using the abstraction over attributes to depict compass direction and altitude range }
         \label{fig:JMAVSim}
\end{figure}

Figure \ref{fig:JMAVSim}a  depicts the abstract state model  obtained by applying range abstraction [324:362:365:385:386:410:411] to the \textit{altitude} of the UAV during its flight from and to its home location at the center point (C) of a circular path. The range of altitude values spanned from 322m to 411m, with an initial value of 0 during initialization. The model includes various states to represent different altitude ranges. 
It is interesting to note that the number of transitions in states (385:386) and (410:411) is approximately equal, reflecting the UAV's two rounds of movement along the circular path.

The combined model in Figure \ref{fig:JMAVSim}b effectively captures both the altitude range and compass direction of the UAV during its flight around the circular path. The compass direction results from a custom abstraction function applied to the latitude and longitude of the UAV. The function maps the latitude and longitude coordinates to one of the eight compass directions relative to the center (C) of the circular path. The home location is represented by the abstract compass value C along with the base altitude range $<$324,  captured by the abstract state (C,$<$324). As the UAV hovers at its home location until takeoff, there is a random movement represented by transitions to and from the state (SW,$<$324), where SW represent South-West direction and state (NE,$<$324), where NE represents North-East direction from the starting state (C,$<$324).  Once the UAV takes off, figure \ref{fig:JMAVSim}b clarifies its ascent to the two altitudes, (385:386) and (410:411), wherein it performs one lap respectively.

By using the abstract values directly to construct the abstract states, the model is able to represent the UAV's movement in a concise and efficient manner. Additionally, the model provides a clear visualization of the two rounds of circular movement at different altitudes, allowing for easy interpretation of the UAV's flight path. Overall, the use of abstraction techniques can greatly simplify the modeling process and make complex systems easier to understand and analyze. Some properties of interest that can be checked during the flight of the UAV are :
\begin{enumerate}
    \item A basic safety property would be to check that the UAV has landed safely at its home location
    (C).  For this purpose, we can check that whenever the altitude {\tt a < 325} the compass direction {\tt dir = "C"}. \\ \\
     {\tt
		G [ (a <= 325 -> dir == "C") ]
	} \\\\
     That the UAV eventually returns to its home location after its intended flight is more akin to a liveness property and can be stated as follows. \\ \\
     {\tt
		G [ (a > 325 \&\& dir != "C") -> F [a <= 325 \&\& dir == "C"] ]
	} \\
    
     \item Another safety property of interest would be that the UAV did not breach its {\it geo-fence}. This property can be checked by applying the abstraction function as the great-circle distance between the current latitude, longitude and its home location coordinates.  This distance
     ({\tt d}) represents the shortest distance over the earth’s surface and can be computed using the {\it Haversine} formula \cite{Haversine}. The {\tt G} property can then be stated as follows, assuming a geo-fence radius of 300m: \\ \\
     {\tt
		G [ d >= 0 \&\& d <= 300 ]
	} \\ 
\end{enumerate}

\subsection{Open Authorization (OAuth) Protocol}

The OAuth 2.0 \cite{OAuth2} is an open-standard protocol that performs secure authorization of resources from an application, whether it be on the web, on a mobile, or on a desktop. This protocol enables an end-user ({\it resource owner}) to share details stored in an application server ({\it resource server}) to a third-party client application ({\it client}) without sharing user credentials to the client application. The protocol enables the end-user to directly authenticate with an authorization server trusted by the client application. The authorization server, upon successful authentication of the resource owner, issues an access token to the client application, using which it can access the protected resource from the resource server. 

\begin{figure}
		\begin{center}
			\includegraphics[height=7in]{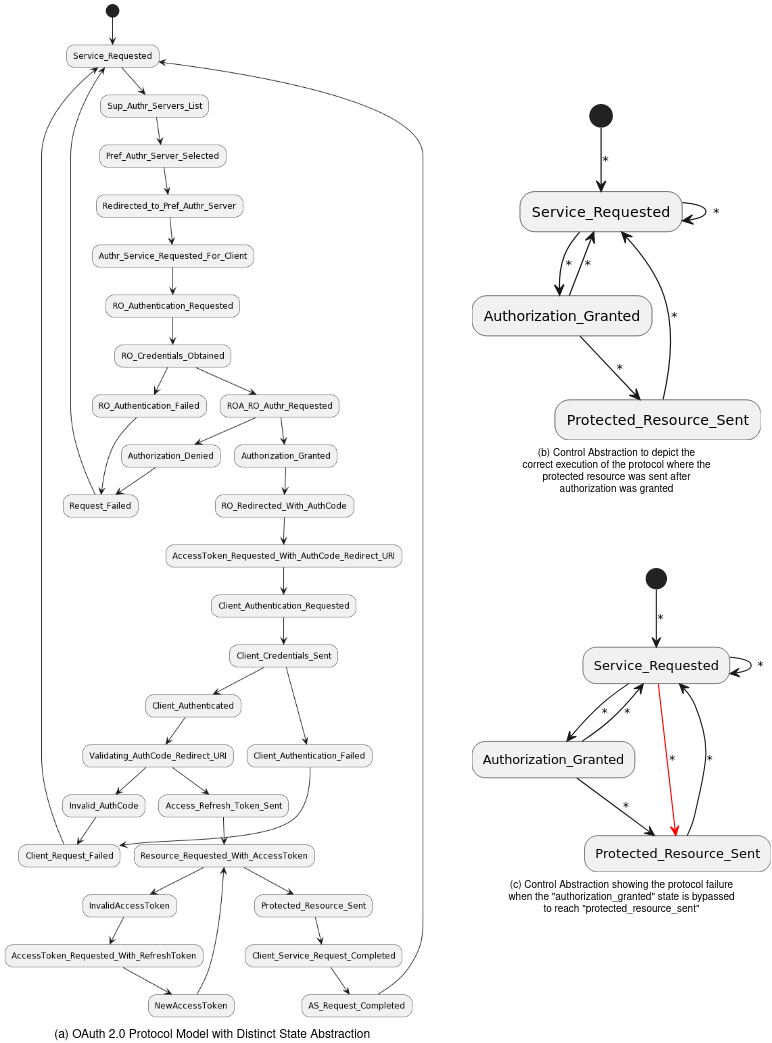}
			\caption{Control Abstraction on the OAuth Model }
			\label{fig:OAuth}
		\end{center}
	\end{figure}

Figure \ref{fig:OAuth}a shows a distinct state model constructed from a run of a Java model of the OAuth protocol that we implemented \cite{JJB:OAuth-2018}.  The linear state model is not shown as it would be very large; however, the most accurate presentation of paths is conveyed by this model rather than the DSM. 
The main property of interest is that every request to access the resource is first authorized. With reference to Figure \ref{fig:OAuth}a the states pertinent to property-checking are stated in the {\tt P} property:
 
\phantom{ PP [ } {\tt P [Service\_Requested $\sim\sim$> Authorization\_Granted $\sim\sim$>}

\phantom{ PPPPI [ } {\tt Protected\_Resource\_Sent]}
\\
\\
The abstract model shown in Figure \ref{fig:OAuth}b is constructed by Algorithm \ref{alg:PathAbstraction} starting from the execution trace and the above {\tt P} property.  
From the abstract model, Algorithm \ref{alg:ASM-PC-Path}
can immediately verify that every path from 
 {\tt Service\_Requested} to {\tt Protected\_Resource\_Sent} passes through {\tt Authorization\_Granted}.  This figure also shows that the resource is not sent in some instances, e.g., when there was a failure before reaching the state {\tt Authorization\_Granted} or before reaching the state {\tt Protected\_Resource\_Sent}.

Suppose the protocol was implemented incorrectly and did not enforce the {\tt P} property. From a run of this incorrect protocol,  Figure \ref{fig:OAuth}c might get constructed. This figure shows that there is a path from state {\tt Service\_Requested} to state {\tt Protected\_Resource\_Sent}  without going through state {\tt Authorization\_Granted}, thereby violating the {\tt P} property.

\subsection{Performance Evaluation}
\label{subsec:performance}

We first present the improvements due to abstraction and then compare our JIV$^2$E online verification system with closely related approaches.

\subsubsection{Improvements due to Abstraction}

Table \ref{tab:Abstraction} presents a comparison of the linear, distinct, and abstract state models with respect to the number of states they each require, for a few applications developed in Java. Classic examples like readers-writers and dining philosophers were chosen for small-scale runs, OAuth protocol for medium-sized runs, and open-source applications for large-scale runs.  We used our custom implementations for the readers-writers, dining-philosophers and OAuth protocol. Since the OAuth protocol involves four parties, a Resource Owner (RO), Resource Server (RS), Authentication Server (AS) and Client Application (C), we were unable to find Java implementation for all these four parties involved in the protocol. Hence, we chose to implement them ourselves to test our system.

Additionally, to evaluate our system on real-world software, we opted for two open-source applications: Apache Tomcat Server, a widely used Java-based application server, and JMAVSim, a Java-based multirotor drone controller. Apache Tomcat Server serves as a good use case for control-oriented models, while JMAVSim serves as an ideal example for data-oriented models. Therefore, we selected these two open-source applications to showcase the effectiveness of our proposed techniques. Apache Tomcat Server was tested for its internal classes and for request processing  with a Java Servlet application deployed on the server. For JMAVSim, various path configurations, including circles, triangles, and random paths, were tested. We had reported the results based on the default circular path available in the controller application (QGroundControl), that enables users to create different paths for the drones.  

\begin{landscape}
	\begin{table*}
		\caption{ Linear (LSM) vs Distinct State (DSM) vs Abstract State Model (ASM)}
		\label{tab:Abstraction}
		\centering
		\begin{tabular*}{600pt}{@{\extracolsep\fill}ccccccc@{\extracolsep\fill}}
			\toprule
			&    &   &  & \multicolumn{3}{@{}c@{}}{\textbf{No of States}}
			\\
			\cmidrule{5-7} 
			{\textbf{Application}} & \makecell{\textbf{Key Attributes} \\ \textbf{(Abbreviations)}}  & \makecell{\textbf{Abstraction} \\ \textbf{Function}}  & \makecell{\textbf{Abstraction} \\ \textbf{Type}} & \textbf{LSM} &  \textbf{DSM} & \textbf{ASM}  \\
			\midrule
			\makecell{Readers-Writers} & \makecell{Database:1.r (r), Database:1.w (w)} & r\textgreater{}0,w=0 & \makecell{Data \\ (Boolean)}& 101 & 7  & 3\\ \\
			
			& \makecell{Database:1.r (r), Database:1.w (w),\\ Database:1.ww (ww)} & ww\textgreater{}0 & \makecell{Data \\ (Boolean)} & 225 & 17  & 11
			\\\\
			\makecell{Dining \\ Philosophers} & \makecell{Philo:1.state (p1), Philo:2.state (p2),\\Philo:3.state (p3),Philo:4.state (p4),\\ Philo:5.state (p5)} & \makecell{p1=E, p2=E, p3=E,\\p4=E, p5=E} & \makecell{Data \\ (Boolean)} & 601 & 50  & 11
			\\\\
			
			JMAVSim & LatLonAlt.alt (a) & \makecell{a={[}324:362:365:385:386:410:411{]}} & \makecell{Data \\ (Range)} & 174923 & 877  & 7\\ \\
			
			& LatLonAlt.hdistance (d) & d \textgreater 299 & \makecell{Data \\ (Boolean)} & 174922 & 301  & 2   \\ \\
			
			& \makecell{LatLonAlt.hdistance (d),\\ LatLonAlt.compassDir (dir)} & d \textgreater 299 & \makecell{Data \\ (Boolean)} & 349843 & 463  & 11 
			\\ \\
			& \makecell{LatLonAlt.alt (a),\\ LatLonAlt.compassDir (dir)} & a \textgreater 325, dir=C& \makecell{Data \\ (Boolean)} & 349843 & 923  & 3 
			\\ \\
			OAuth & Scheduler.scheduled (s) & \makecell{  s=Service\_Requested, \\ s=Authorization\_Granted, \\ s=Protected\_Resource\_Sent } & \makecell{Control \\ (Path)} & 10499 & 29  & 3  
			\\ \\
			\makecell{Apache Tomcat \\ Server} & cfi.currentMethod (c) & \makecell { c=Bootstrap.main, \\ c=Bootstrap.load, \\ c=Catalina.load  } & \makecell{Control \\ (Path)} & 8149700 & 1429 & 3 \\ \\
			\makecell{Apache Tomcat \\ Servlet Application} & cfi.currentMethod (c) & \makecell { c = EntryServlet.init, \\c = EntryServlet.doGet,\\ c = EntryServlet.destroy} & \makecell{Control \\ (Path)} & 85270 & 45 & 3 \\ 
			\bottomrule
		\end{tabular*}
	\end{table*}
\end{landscape}

\begin{landscape}
	\begin{table*}
		\caption{ Model construction, Model verification and Total Time taken for verification on Linear, Distinct and Abstract State Models }
		\label{tab:AbstractionTime}
		\centering
		\begin{tabular*}{650pt}{@{\extracolsep\fill}ccccccccccc@{\extracolsep\fill}}
			
			\toprule
			
			&   & \multicolumn{3}{@{}c@{}}{\textbf{ Model Cons. Time (ms)}} & \multicolumn{3}{@{}c@{}}{\textbf{ Model Verif. Time (ms)}} & \multicolumn{3}{@{}c@{}}{\textbf{Total Time (ms)}}
			\\
			
			\cmidrule{3-5}  \cmidrule{6-8} \cmidrule{9-11}
			
			\makecell{\textbf{Application} / \\ \textbf{(Abstraction Function)}}  & \makecell{\textbf{Property} \\ \textbf{Verified}} & \textbf{LSM} &  \textbf{DSM} & \textbf{ASM} & \textbf{LSM} &  \textbf{DSM} & \textbf{ASM}  & \textbf{LSM} &  \textbf{DSM} & \textbf{ASM}   \\
			
			\midrule
			
			\makecell{Readers-Writers  /\\  (r\textgreater{}0, w=0) } & \makecell{ G[ r\textgreater{}0 $\rightarrow$  w=0 ]}& 0.80 & 0.59  & 0.46 & 0.36 & 0.08 & 0.05 &1.16 &0.67 &0.51\\ \\

			\makecell{ Dining Philosophers / \\   (p1=E, p2=E, \\ p3=E,p4=E, p5=E)  } & \makecell{G[(p1=E $\rightarrow$ p2!=E) \&\& \\ (p2=E $\rightarrow$ p3!=E) \&\& \\ (p3=E $\rightarrow$ p4!=E) \&\& \\ (p4=E $\rightarrow$ p5!=E) \&\& \\ (p5=E $\rightarrow$ p1!=E)]} & 10.24 & 6.68  & 11.21 & 2.11 & 0.36 & 0.45 & 12.35&7.04 &11.66
			\\ \\
			
			\makecell{ JMAVSim / \\  (a\textgreater325, dir=C)} & \makecell{G[ a$<=$325 $\rightarrow$ dir=C ]} & 884& 899 & 1453 & 163 & 6.65 & 1.57 & 1047 & 906 & 1455\\ \\
			
			\makecell{ JMAVSim / \\ (d \textgreater 299) }& \makecell{G[ d$<=$299 ]} & 568 & 868  & 700  & 59.25 & 2.65 & 0.40 & 627 & 871& 700\\ \\

			\makecell{ OAuth /\\ (s=Service\_Requested, \\ s=Authorization\_Granted, \\ s=Protected\_Resource\_Sent)} & \makecell{P[s=Service\_Requested \\$\sim\sim>$ \\s=Authorization\_Granted \\$\sim\sim>$ \\s=Protected\_Resource\_Sent]} & 0.80 & 0.63
			& 0.63 & 0.26 & - 
			& 0.07  & 1.06 & - & 0.70
			\\ \\
			
			\makecell{Apache Tomcat  Server /\\ (c=Bootstrap.main, \\ c=Bootstrap.load, \\ c=Catalina.load) } & \makecell{P[c=Bootstrap.main \\$\sim\sim>$ \\c=Bootstrap.load \\$\sim\sim>$ \\c=Catalina.load]} & 22568 & 20629
			& 7601 & 5571 & - 
			& 0.25 & 28139& - & 7601\\ \\
			
			\makecell{Apache Tomcat \\ Servlet Application /\\ (c = EntryServlet.init, \\c = EntryServlet.doGet,\\ c = EntryServlet.destroy) } & \makecell{P[c=EntryServlet.init \\$\sim\sim>$ \\c=EntryServlet.doGet \\$\sim\sim>$ \\c=EntryServlet.destroy]} & 137 &  77.34
			& 44.90 & 32.32 & - 
			& 1.36 & 169.32 & - & 46.26\\ 
			\bottomrule
		\end{tabular*}
	\end{table*}
\end{landscape}

The Linear State Model (LSM) requires the highest number of states, as it creates a new state for every \textit{Field Write} event occurring for the selected {key attribute(s)} in the execution trace. However, as noted earlier, it is the most accurate model to represent the execution of the application. The Distinct State Model (DSM) eliminates state duplication which can arise in an LSM by creating a new state only for each unique combination of values of the key attributes. 

In assessing the benefit of model abstraction, we note that comparison against the Linear State Model could give an inflated view of improvement because the number of states in the LSM can be arbitrarily increased by increasing the length of the execution trace.
Thus the benefit of model abstraction is better assessed by comparing the number of states
in the Distinct State Model (DSM) with the corresponding Abstract State Model (ASM).  The extent of improvement, of course,
depends upon the specific abstraction used.  Table \ref{tab:Abstraction} shows our
results for abstractions that are natural for the problems at hand.

Readers-Writers is a small example and the figures here are not typical of larger examples.  However, it is included here for completeness.
For the Dining Philosophers example, we can show that the maximum number of possible states in the discrete state model is 81 (regardless of the length of the execution trace), assuming that the state thinking ({\tt T}) means that the philosopher has no forks in hand, hungry ({\tt H}) means that the philosopher has one fork in hand, and eating ({\tt E}) means that he has both forks.  For the specific run reported in the table, 50 out of the 81 possible states were encountered; however, all 11 abstract states were present.While the maximum possible  improvement due to model abstraction is a factor of 7.36 (81/11), the observed improvement for the run reported is a factor of 4.54 (50/11).

The OAuth Protocol table entry illustrates the typical improvement from path abstraction.  The distinct state model has 29 states while path abstraction brings that down to just three states in the abstract model -- for service requested, authorization granted, and resource sent. Thus, the improvement is a factor of 9.66 (29/3).

The table entries for the JMAVSim application illustrate the improvement due to multi-valued (range) and boolean abstraction. The distinct state model has between 300-900 states, while abstraction brings that down by a factor of 100.
The table also has entries for performance improvements of model abstraction in the Apache Tomcat Server which was discussed extensively in our earlier paper \cite{JJJS:FSM-SPE-2021} and whose details are reported on our website.  Here, path properties are more common and the improvements due to abstraction are a factor of 100 on the average.

Table \ref{tab:AbstractionTime} presents a comparison of the time taken for model construction, property verification and the total time for linear, distinct, and abstract state models for the applications depicted in Table \ref{tab:Abstraction}. All timings were obtained by running the applications on a 2.7 GHz Intel Core i5 processor with 8 GB RAM. Table \ref{tab:AbstractionTime}  shows the abstractions applied on the key attributes to construct the abstract models, followed by the property verified. While the Readers-Writers, Dining Philosophers and JMAVSim demonstrate performance of data abstraction, OAuth and Apache Tomcat server and Apache Tomcat Servlet application demonstrate the performance of control(path) abstraction. We can see that the model construction time is slightly higher for data-oriented ASM models compared with their respective LSM models. For control-oriented (path) abstraction, ASM models require less time than their respective LSM models. In both cases, property verification time is a factor of 10-20 less for ASM models, for smaller runs, and around 100 less for large runs compared with the corresponding LSM counterparts. With respect to the total time, we can observe that the ASM-based verification is competitive with LSM for small data-oriented applications, whereas the LSM outperforms ASM for large data-oriented applications such as JMAVSim. For the control-oriented properties, ASM-based verification outperforms LSM-based verification. As noted earlier, since the path properties cannot be tested correctly on DSM, we have not carried out verification of path properties under DSM. 

\subsubsection{Comparison with other Approaches}

The JIV$^2$E verification system supports two modes of verification: {\it offline} and {\it online}. Offline verification is suitable for retrospective analysis of the system. Here, states are {\it not} created as the events are received and verification is also not carried until the end of the trace. The events are generally written to a file during program execution for later offline analysis. In contrast, in online verification,  as the events are received from program execution, the states are created on-the-fly and verification is also carried out.  

The time complexity of property checking on the Linear State Model(LSM) is $O(n)$, where $n$ represents the length of the input trace and the formula size is considered as a constant. 
The memory required for LSM is $O(n.m)$ 
where $m$ is the number of key attributes chosen to create the runtime model.
In the JIV$^2$E online verification system, we explicitly represent the states of execution which aids visualization and property checking.   These states are 
abstract states in practice unlike the monitoring systems discussed below.

In comparison with other closely related approaches, which are  mainly monitoring systems such as MaC \cite{IBKKSV:MaC-1998, KVKIS:Java-MaC-FMSD-2004}, JPAX\cite{HG:RV-JPAX-MonitorSynthesis-2002,HG:RV-JPAX-2004} and MOP \cite{FG:MOP-2007, MJGG:MOP-2012}, their property checking on the concrete trace comes closest to online property checking on our Linear State Model. These monitoring systems also maintain an internal state to evaluate properties,
but they do not use the concept of key attributes, as in our approach, nor do they incur the
same memory cost.

The {event recognizer} of the MaC architecture \cite{IBKKSV:MaC-1998, KVKIS:Java-MaC-FMSD-2004} maintains a \textit{variable/method} table to track the names and the current values of monitored variables and methods, and a \textit{event/condition} table to track the names and references to the events/condition trees. These events/condition trees are reevaluated every time a new event is received from the monitored system through MaC framework's {filter} module. 
 
JPAX\cite{HG:RV-JPAX-MonitorSynthesis-2002} has in addition to an internal state, two arrays, \textit{pre} and \textit{now}, for tracking values
in the previous and current state respectively, for every formula to be monitored. The internal state is updated through an \textit{update} function whenever an event is received. 
In MOP \cite{MJGG:MOP-2012}, the finite state monitors use internal state representations to keep track of the system state and update them as the events are received. 

In JIV$^2$E the \textit{key attributes} serve as a basis for constructing the runtime model and the properties are evaluated on the states of this model. 
Unlike other systems, JIV$^2$E supports property checking on abstract models, wherever suitable abstractions are feasible.  
This reduces the load on the monitor because it needs to verify properties only on newly added abstract states. 
For abstract models, abstract states are created instead of concrete states and the properties are immediately verified as each abstract state is created. Hence property checking is performed once per abstract state rather than once per event received as is done in the linear state model or other monitoring frameworks \cite{IBKKSV:MaC-1998}.

Regarding the time taken for model construction, the algorithms given in section \ref{sec:abstraction} all require one traversal through the execution trace.  
Knowledge of the abstraction criteria allows us to consolidate model construction, i.e., we can construct the distinct and all abstract models with just one pass through the trace.
As noted earlier, the use of an incorrect abstraction can cause property checking to fail, necessitating re-traversal of the execution trace to re-construct the
abstract model.  In practice this happens when one is uncertain
with the choice of abstraction or one is still gaining
an in-depth understanding about the application being checked. The number of abstract states is related to the sizes of the abstract domains.  In practice, this is not a large number, as  illustrated by Table \ref{tab:Abstraction}, 
and hence property checking on abstract models can be carried out very efficiently.

\section{Conclusions and Further Work}
\label{sec:conclusion}

The focus of this paper is on run-time model abstraction and its benefit in obtaining
smaller models and greater efficiency in run-time verification. 
Property preservation is an important criterion during abstraction, because it ensures that verification on the smaller, abstract model carries over to the larger, concrete model. 
It should be noted that abstraction is not always needed and property checking can be carried out directly on the concrete model when it is not large.
Also, not all abstractions support property preservation; over-abstraction or an inappropriate abstraction may hinder property verification and under-abstraction may result in unnecessarily large models.  This topic has been studied extensively at design-time \cite{Sifakis:PropPreservingHomomorphisms-1982}, \cite{CGL:MCAbst-Clarke-1994},\cite{BLR-Ten-SLAM}, \cite{Dwyer:Bandera-Abstraction-2001}. 
The contributions of this paper are on abstract finite state models derived from execution traces of Java programs and run-time verification of correctness properties on the abstract models.  
 
An important insight from our work is that the lower levels of a software require data-oriented models and associated abstractions (boolean- or multi-valued) whereas the higher levels of a software require control-oriented models and associated abstractions (code-structure- or path-based).
Also, the size of the software is not always related to the length of an execution run or the size of the generated model, as large software can result in small runs/models and small-to-medium software can result in long runs/models.

We illustrated how abstractions can be chosen to be compatible with the property to be verified and we provided an informal justification of the correctness of boolean-, multi-valued, and path-abstractions.

As the focus of our work is on runtime verification, our analysis is done on a finite execution trace.    
We essentially focus on state-based or path-based that can be checked within the limits of a finite trace. 
As our examples are programs such as servers and controllers with a cyclic operation, a single run typically encompasses many (though not all) execution scenarios.
The paper illustrated our techniques on small, classic examples from the concurrency literature as well as larger well-known case studies -- the OAuth 2.0 Protocol and a Multi-rotor Drone Controller.   

To cater to long executions, we employ a custom byte-code instrumentation technique that 
facilitates efficient offline as well as online verification of properties. 
In online verification, abstract states are created as events are generated and property checking
is carried out immediately for state-based properties.  Path-based properties cannot always be checked immediately because they might depend upon future states not already encountered.
Our algorithms for model abstraction can bypass the construction of a concrete model
and directly yield concise abstract models.  We have incorporated these techniques in our evolving prototype, JIV$^2$E, which has been developed over the past 15 years.  
Our experimental results shown in Table 
\ref{tab:AbstractionTime} justify the conclusion that model abstraction is a 
practical and efficient technique for runtime verification.

Program monitoring with dynamic property checking are practical techniques for many applications. 
However, these applications are often developed in C or C++ and hence our techniques will have to be adapted to apply to these contexts.  We are currently exploring this topic, the key problem here being the generation of execution events, whether online or offline.  These languages support compilation with a debug option so that
source-level information is available at execution time.  
When the compiled codes for these languages use common debug formats we can take advantage of the APIs of a tool such as Dyninst \cite{Dyninst-URL} 
for adding instrumentation code that could
generate JIV$^2$E events.  These in turn can be input to our model construction and runtime verification system. 

As part of our future work, we also plan to enhance  JIV$^2$E  to support animated and interactive visualizations that can be used to comprehend the functioning of unknown systems.  We also plan to interactively try different abstractions on the fly that aid in deducing properties that hold on the system under test.  Another line of work will include evolving the architecture to support distributed monitoring, thereby supporting parallel verification of properties on the executing system.
 
 \section*{\refname}
\bibliography{mybibfile}

\begin{thebibliography}{10}
\expandafter\ifx\csname url\endcsname\relax
  \def\url#1{\texttt{#1}}\fi
\expandafter\ifx\csname urlprefix\endcsname\relax\def\urlprefix{URL }\fi
\expandafter\ifx\csname href\endcsname\relax
  \def\href#1#2{#2} \def\path#1{#1}\fi

\bibitem{GJ:JIVE-2005}
P.~Gestwicki, B.~Jayaraman, {Methodology and Architecture of \textup{JIVE}},
  in: Proc. ACM Symp. on Software Visualization, 2005, pp. 95--104.

\bibitem{CJ:ETX-2007}
J.~K. Czyz, B.~Jayaraman, {Declarative and Visual Debugging in Eclipse}, in:
  Proc. ACM OOPSLA Works. on Eclipse Technology Exchange, {Association for
  Computing Machinery}, 2007, pp. 31--35.

\bibitem{LBDSwaminathan:RV-JIVE-2016}
L.~Ziarek, B.~Jayaraman, D.~Lessa, S.~Jayaraman, {Runtime Visualization and
  Verification in {JIVE}}, in: Y.~Falcone, C.~S{\'{a}}nchez (Eds.), 16th Intl.
  Conf. on Runtime Verification, Vol. LNCS 10012, Springer, 2016, pp. 493--497.

\bibitem{JLSwaminathan:SPE-2017}
S.~Jayaraman, B.~Jayaraman, D.~Lessa, {Compact Visualization of Java Program
  Execution}, Software: Pract. and Experience 47~(2) (2017) 163--191.

\bibitem{JJJS:FSM-SPE-2021}
K.~P. Jevitha, S.~Jayaraman, B.~Jayaraman, M.~Sethumadhavan, {Finite-state
  Model Extraction and Visualization from Java Program Execution}, Software:
  Pract. and Experience 51~(2) (2021) 409--437.

\bibitem{JBR:UML-1999}
I.~Jacobson, G.~Booch, J.~Rumbaugh, {The Unified Software Development Process},
  Addison-Wesley, Boston, MA, 1999.

\bibitem{Sifakis:PropPreservingHomomorphisms-1982}
{Sifaks, Joseph}, {Property-Preserving Homomorphisms of Transition Systems},
  in: {E. Clarke and D. Kozen, editors, Works. on Logics of Programs, LNCS
  164}, Springer-Verlag, 1983.

\bibitem{Clarke:AutomaticVerification-FSCS-TLS-1986}
E.~Clarke, E.~Emerson, A.~Sistla, {Automatic Verification of Finite-State
  Concurrent Systems Using Temporal Logic Specifications}, ACM Trans. on
  Programming Languages and Systems 8~(2) (1986) 244--263.

\bibitem{Harel:StateCharts-1987}
D.~Harel, {Statecharts: A Visual Formalism for Complex Systems}, Science of
  Computer Programming 8~(3) (1987) 231--274.

\bibitem{Clarke:ModelChecking-1997}
E.~M. Clarke, {Model Checking}, in: S.~Ramesh, G.~Sivakumar (Eds.), Foundations
  of Software Technology and Theoretical Computer Science, Springer, 1997.

\bibitem{HG:RV-2001-ase}
K.~Havelund, G.~Rosu, {Monitoring Programs using Rewriting}, in: Proceedings of
  16th IEEE International Conference. Automated Software Engineering (ASE'01),
  IEEE, 2001, pp. 135--143.

\bibitem{MC:RV-Brief-2009}
M.~Leucker, C.~Schallhart, {A Brief Account of Runtime Verification}, The
  Journal of Logic and Algebraic Programming 78~(5) (2009) 293--303.

\bibitem{FHR:RV-tutorial-2013}
Y.~Falcone, K.~Havelund, G.~Reger, {A Tutorial on Runtime Verification},
  Engineering Dependable Software Systems (2013) 141--175.

\bibitem{HSZ-Sistla:MC+RV-2014}
{T.L. Hinrichs, A.P. Sistla, L.D. Zuck}, {Model Check What You Can, Runtime
  Verify the Rest}, in: {HOWARD-60: A Festschrift on the Occasion of Howard
  Barringer's 60th Birthday}, Vol.~42, 2014, pp. 234--244.

\bibitem{LGSBB:PropPreservingAbstractions-1995}
C.~Loiseaux, S.~Graf, J.~Sifakis, A.~Bouajjani, S.~Bensalem, {Property
  Preserving Abstractions for the Verification of Concurrent Systems}, Formal
  Methods in System Design 6~(1) (1995) 11–44.

\bibitem{JMAVSim1}
{JMAVSim}, 2023, \url{ https://docs.px4.io/main/en/simulation/jmavsim.html}.

\bibitem{OAuth2}
D.~Hardt, {The OAuth 2.0 Authorization Framework}, RFC 6749 (October 2012).

\bibitem{tomcatarch}
\href{https://tomcat.apache.org/tomcat-9.0-doc/architecture/index.html}{{The
  Apache Software Foundation}}, {Apache Tomcat 9 Architecture, 2019}.
\newline\urlprefix\url{https://tomcat.apache.org/tomcat-9.0-doc/architecture/index.html}

\bibitem{CGL:MCAbst-Clarke-1994}
E.~M. Clarke, O.~Grumberg, D.~E. Long, {Model Checking and Abstraction}, ACM
  Trans. on Programming Languages and Systems 16~(5) (1994) 1512–1542.

\bibitem{BMMR:PredicateAbstractionCPgms-2001}
T.~Ball, R.~Majumdar, T.~Millstein, S.~Rajamani, {Automatic Predicate
  Abstraction of C Programs}, in: Proc. ACM Conf. on Programming Language
  Design and Implementation, 2001, p. 203–213.

\bibitem{BLR-Ten-SLAM}
T.~Ball, V.~Levin, S.~Rajamani, {A Decade of Software Model Checking with
  SLAM}, {Comm. of the ACM} 54~(7) (2011) 68--76.

\bibitem{Dwyer:Bandera-Abstraction-2001}
M.~B. Dwyer, J.~Hatcliff, R.~Joehanes, S.~Laubach, C.~S. P\u{a}s\u{a}reanu,
  H.~Zheng, W.~Visser, {Tool-Supported Program Abstraction for Finite-State
  Verification}, in: Proc. 23rd IEEE Intl. Conf. on Software Engineering, 2001,
  pp. 177--–187.

\bibitem{CC:AbstractInterpretation-1977}
P.~Cousot, R.~Cousot, {Abstract Interpretation: A Unified Lattice Model for
  Static Analysis of Programs by Construction or Approximation of Fixpoints},
  in: Proc. 4th ACM Symp. on Principles of Programming Languages, 1977, pp.
  238--252.

\bibitem{CC:AbstractInterpretation-1979}
P.~Cousot, R.~Cousot, {Systematic Design of Program Analysis Frameworks}, 1979.

\bibitem{DLWZ:MiningObjectBehaviourADABU-2006}
V.~Dallmeier, C.~Lindig, A.~Wasylkowski, A.~Zeller,
  \href{https://doi.org/10.1145/1138912.1138918}{Mining object behavior with
  adabu}, in: Proceedings of the 2006 International Workshop on Dynamic Systems
  Analysis, WODA '06, Association for Computing Machinery, New York, NY, USA,
  2006, p. 17–24.
\newblock \href {http://dx.doi.org/10.1145/1138912.1138918}
  {\path{doi:10.1145/1138912.1138918}}.
\newline\urlprefix\url{https://doi.org/10.1145/1138912.1138918}

\bibitem{MTR:StateBasedTesting-AJAX-2008}
A.~Marchetto, P.~Tonella, F.~Ricca, {State-Based Testing of Ajax Web
  Applications}, in: 2008 1st International Conference on Software Testing,
  Verification, and Validation, 2008, pp. 121--130.
\newblock \href {http://dx.doi.org/10.1109/ICST.2008.22}
  {\path{doi:10.1109/ICST.2008.22}}.

\bibitem{MMNTB:Revolution-AutomaticEvolutionMinedSpecs-2012}
L.~Mariani, A.~Marchetto, C.~D. Nguyen, P.~Tonella, A.~Baars, {Revolution:
  Automatic Evolution of Mined Specifications}, in: 2012 IEEE 23rd
  International Symposium on Software Reliability Engineering, 2012, pp.
  241--250.
\newblock \href {http://dx.doi.org/10.1109/ISSRE.2012.14}
  {\path{doi:10.1109/ISSRE.2012.14}}.

\bibitem{ODDL:ControlledBurstRecording-CBR-2020}
O.~Cornejo, D.~Briola, D.~Micucci, L.~Mariani,
  \href{https://api.semanticscholar.org/CorpusID:211031892}{{CBR: Controlled
  Burst Recording}} (2020) 243--253.
\newline\urlprefix\url{https://api.semanticscholar.org/CorpusID:211031892}

\bibitem{OCL}
{Object Constraint Language - Version 2.4},
  \url{https://www.omg.org/spec/OCL/}.

\bibitem{PM-2003-OCL}
P.~Ziemann, M.~Gogolla, {OCL Extended with Temporal Logic}, in: Perspectives of
  System Informatics, Springer Berlin Heidelberg, Berlin, Heidelberg, 2003, pp.
  351--357.

\bibitem{KT-2012}
{B. Kanso and S. Taha}, {Temporal constraint support for OCL}, in: {Intl. Conf.
  on Software Language Engineering}, {Springer, Berlin, Heidelberg}, 2012, pp.
  {83--103}.

\bibitem{IBKKSV:MaC-1998}
{I. Lee, H. Ben-Abdallah, S. Kannan, M. Kim, O. Sokolsky, M. Viswanathan}, {A
  Monitoring and Checking Framework for Run-time Correctness Assurance}, in:
  {Proc. Korea-U.S. Tech. Conf. on Strategic Technologies, Vienna}, 1998.

\bibitem{KVKIS:Java-MaC-FMSD-2004}
M.~Kim, M.~Viswanathan, S.~Kannan, I.~Lee, O.~Sokolsky, {Java-MaC: A Run-Time
  Assurance Approach for Java Programs}, {Formal methods in System Design} 24
  (2004) 129--155.

\bibitem{HG:RV-JPAX-2001}
K.~Havelund, G.~Roşu, {Monitoring Java Programs with Java PathExplorer},
  Electronic Notes in Theoretical Computer Science 55~(2) (2001) 200--217,
  {Runtime Verification (in connection with CAV 2001)}.

\bibitem{HG:RV-JPAX-2004}
K.~Havelund, G.~Ro{\c{s}}u, {An Overview of the Runtime Verification Tool Java
  PathExplorer}, Formal Methods in System Design 24~(2) (2004) 189--215.

\bibitem{BGHS:EAGLE-2004}
H.~Barringer, A.~Goldberg, K.~Havelund, K.~Sen, {Rule-based Runtime
  Verification}, in: International Workshop on Verification, Model Checking,
  and Abstract Interpretation, Springer, 2004, pp. 44--57.

\bibitem{MarceloHavelund:RV-Java-HAWK-2005}
M.~D'Amorim, K.~Havelund, {Event-Based Runtime Verification of Java Programs},
  ACM SIGSOFT Software Engineering Notes 30~(4) (2005) 1–7.

\bibitem{Allan:TRACEMATCHES-AspectJ-2005}
C.~Allan, P.~Avgustinov, A.~S. Christensen, L.~Hendren, S.~Kuzins,
  O.~Lhot{\'a}k, O.~De~Moor, D.~Sereni, G.~Sittampalam, J.~Tibble, {Adding
  Trace Matching With Free Variables to AspectJ}, ACM SIGPLAN Notices 40~(10)
  (2005) 345--364.

\bibitem{StolzBodden:JLO-AspectJ-2006}
V.~Stolz, E.~Bodden, {Temporal Assertions Using AspectJ}, Electronic notes in
  theoretical computer science 144~(4) (2006) 109--124.

\bibitem{CG:MOP-2003}
F.~Chen, G.~Roşu, {Towards Monitoring-Oriented Programming: A Paradigm
  Combining Specification and Implementation}, Electronic Notes in Theoretical
  Computer Science 89~(2) (2003) 108--127.

\bibitem{CMG:MOP-2004}
F.~Chen, M.~D'Amorim, G.~Ro{\c{s}}u, A formal monitoring-based framework for
  software development and analysis, in: Formal Methods and Software
  Engineering, Springer Berlin Heidelberg, 2004.

\bibitem{FG:MOP-2007}
F.~Chen, G.~Ro\c{s}u, {MOP: An Efficient and Generic Runtime Verification
  Framework}, SIGPLAN Not. 42~(10) (2007) 569–588.

\bibitem{CG:MOP-2009}
F.~Chen, G.~Ro{\c{s}}u, {Parametric Trace Slicing and Monitoring}, in:
  International Conference on Tools and Algorithms for the Construction and
  Analysis of Systems, Springer, 2009, pp. 246--261.

\bibitem{MJGG:MOP-2012}
P.~O. Meredith, D.~Jin, D.~Griffith, F.~Chen, G.~Ro{\c{s}}u, {An Overview of
  the MOP Runtime Verification Framework}, International Journal on Software
  Tools for Technology Transfer 14~(3) (2012) 249--289.

\bibitem{Kiczales:AOP-1997}
G.~Kiczales, J.~Lamping, A.~Mendhekar, C.~Maeda, C.~Lopes, J.-M. Loingtier,
  J.~Irwin, Aspect-oriented programming, in: M.~Ak{\c{s}}it, S.~Matsuoka
  (Eds.), Proc. ECOOP: Object-Oriented Programming, 1997, pp. 220--242.

\bibitem{Binder:DiSL-dynpgmAnys-SCP-2015}
L.~Marek, Y.~Zheng, D.~Ansaloni, L.~Bulej, A.~Sarimbekov, W.~Binder, P.~Tuma,
  {Introduction to Dynamic Program Analysis with DiSL}, Science of Computer
  Programming 98~(P1) (2015) 100–115.

\bibitem{GGANJ-Disl-2023}
G.~Saravanan, G.~Subramani, P.~N. S.~S. Akshay, N.~Kanigolla, K.~P. Jevitha,
  {Run-time Control Flow Model Extraction of Java Applications}, in: Emerging
  Research in Computing, Information, Communication and Applications, Springer
  Nature Singapore, 2023, pp. 803--816.

\bibitem{DV:LTLf-2013}
G.~De~Giacomo, M.~Y. Vardi, {Linear Temporal Logic and Linear Dynamic Logic on
  Finite Traces}, in: Proceedings of the Twenty-Third International Joint
  Conference on Artificial Intelligence, IJCAI '13, AAAI Press, 2013, p.
  854–860.

\bibitem{Dijkstra:DP-1971}
E.~W. Dijkstra, {Hierarchical Ordering of Sequential Processes}, Acta
  Informatica 1~(2) (1971) 115--138.

\bibitem{CHP:RW-1971}
P.~J. Courtois, F.~Heymans, D.~L. Parnas, {Concurrent Control with
  “Readers” and “Writers”}, Comm. of the ACM 14~(10) (1971) 667–668.

\bibitem{javalanginstrumentJavaSE8}
\href{https://docs.oracle.com/javase/8/docs/api/java/lang/instrument/package-summary.html}{{java.lang.instrument
  (Java Platform SE 8 )}}.
\newline\urlprefix\url{https://docs.oracle.com/javase/8/docs/api/java/lang/instrument/package-summary.html}

\bibitem{ASM2002}
E.~Bruneton, R.~Lenglet, T.~Coupaye, {\textup{ASM}: A Code Manipulation Tool to
  Implement Adaptable Systems}, Adaptable and Extensible Component Systems
  30~(19).

\bibitem{ASM2007}
E.~Kuleshov, {Using \textup{ASM} Framework to Implement Common Bytecode
  Transformation Patterns}, in: Proc. 6th ACM Aspect Oriented Software
  Development, 2007.

\bibitem{JMAVSim2}
{JMAVSim}, 2023. \url{ https://github.com/PX4/jMAVSim}.

\bibitem{PX4conf}
L.~Meier, D.~Honegger, M.~Pollefeys, {PX4: A Node-Based Multithreaded Open
  Source Robotics Framework for Deeply Embedded Platforms}, in: 2015 IEEE
  International Conference on Robotics and Automation (ICRA), 2015, pp.
  6235--6240.
\newblock \href {http://dx.doi.org/10.1109/ICRA.2015.7140074}
  {\path{doi:10.1109/ICRA.2015.7140074}}.

\bibitem{PX4}
{Open Source for Drones-PX4 Pro Open Source Autopilot}, {PX4 Dev Team, 2023.
  \url{ https://px4.io}}.

\bibitem{Haversine}
{W. Gellert, S. Gottwald, M. Hellwich, H. K\"astner, H. K\"ustner}, {The VNR
  Concise Encyclopedia of Mathematics, 2nd edition, chapter 12}, {Van Nostrand
  Reinhold: New York}, New York, NY, USA, 1989.

\bibitem{JJB:OAuth-2018}
{Jayasri, K.S., Jevitha, K.P. and Jayaraman, B.}, {Verification of OAuth 2.0
  using Uppaal}, in: Proc. Computer Society of India, Springer, 2018, pp.
  58--67.

\bibitem{HG:RV-JPAX-MonitorSynthesis-2002}
K.~Havelund, G.~Ro{\c{s}}u, {Synthesizing Monitors for Safety Properties}, in:
  J.-P. Katoen, P.~Stevens (Eds.), {Tools and Algorithms for the Construction
  and Analysis of Systems}, Springer Berlin Heidelberg, 2002, pp. 342--356.

\bibitem{Dyninst-URL}
{Dyninst API - University of Maryland and University of Wisconsin},
  \url{https://dyninst.org/dyninst}.

\end{thebibliography}
\end{document}